\documentclass[prb,twocolumn,amsmath,amssymb,floatfix]{revtex4}
\usepackage{graphicx}

\newcommand{\be}{\begin{eqnarray}}
\newcommand{\ee}{\end{eqnarray}}

\newcommand{\bfr}{{\bf r}}
\newcommand{\bfq}{{\bf q}}

\newcommand{\bfk}{{\bf k}}

\newcommand{\bfn}{{\bf n}}

\newcommand{\bfR}{{\bf R}}
\newcommand{\bfG}{{\bf G}}

\newcommand{\tlg}{\tilde{g}}
\newcommand{\tlc}{\tilde{c}}
\newcommand{\tlm}{\tilde{m}}

\newcommand{\wbe}{\begin{widetext}}
\newcommand{\wee}{\end{widetext}}
\newcommand{\oncite}{\onlinecite}

\begin{document}

\title{Thermal and quantum noncondensate particles 
near the superfluid to Mott insulator transition}

\author{Daw-Wei Wang}

\affiliation{Physics Department, 
National Tsing-Hua University, Hsinchu, Taiwan, ROC
\\
Physics Division, National Center for Theoretical Science, Hsinchu, 
Taiwan, ROC
}

\date{\today}

\begin{abstract}
We investigate the finite temperature momentum distribution 
of bosonic noncondensate particles inside a 3D optical lattice 
near the superfluid to Mott insulator transition point, treating
the quantum fluctuation and thermal fluctuation effects on equal
footing. We explicitly address the different momentum ($\bfq$) dependence
of quasi-particle distribution resulted from thermal and 
quantum origins: 
the former scales as $|\bfq|^{-2}$ and hence is dominant in the
small momentum region, while the later scales as $|\bfq|^{-1}$
and hence dominant in the large momentum limit.
Analytic and semi-analytic results are derived, providing a unique 
method to determine various properties inside the optical lattice, including
temperature, condensate density, coherent length and/or single particle gap etc.
Our results also agree with the scaling theory of a quantum 
$XY$ model near the transition point. Experimental implication 
of the TOF measurement is also discussed.
\end{abstract}


\maketitle
\section{Introduction:}
\label{introduction}

The experimental realization of superfluid (SF) to Mott insulator (MI)
transition of ultracold atoms [\oncite{SF_MI_bloch,SF_MI_others}] in 
an optical lattice has open a new area of strongly correlated physics 
and lead to many applications to other fields [\oncite{review}].
It is generally believed that when the size of the expanding atom 
cloud is much larger 
than the initial size (i.e. a long-time flight after switching off the
trapping potential) and if the interaction effect during the
expansion can be neglected [\oncite{interaction_TWA}], 
the TOF image can be interpreted as 
a momentum distribution function of the initial atom cloud inside
the optical lattice. As a result, the sharp interference peaks at zero
momentum and Bragg momentum can be understood as a signature
of a Bose-Einstein condensation (or superfluidity) while the wider
and smaller hump around the peak is then attributed to the 
non-condensate particles inside the optical lattice. 
Different from the superfluidity measurement of condensate via 
vortices in a single parabolic confinement potential 
[\oncite{Nature_fermion_pair_MIT}], 
the time-of-flight (TOF) absorption image (i.e.
the "bi-modal" structure with a "sharp" interference peak in the TOF
image) is so far the 
only experimental measurement to determine the superfluid/condensate
density inside the optical lattice 
[\oncite{Nature_fermion_pair_MIT,B_F_ETH}]. 

However, behind the scenario of "bi-modal" structure of the interference
peak, there is an important assumption: only two length 
scales (or momentum scales) are relevant in the long wavelength limit 
and their difference can be distinguishable [\oncite{Stanford}]: 
one is associate with the size of condensate and the other is associate with the
thermal wavelength of non-condensate particles.
In fact, from experiment point of view, only after the later
part (non-condensate particles) has been identified and subtracted 
from the TOF image spectrum, one can study
the properties (i.e. the sharpness, width, or condensate fraction) 
of the former (condensate) part, which is the key player to 
determine the many-body
phase diagram and has been extensively discussed in recent theoretical 
and experimental groups [\oncite{Jason_sharp,QMC_Ohio,
temp_lattice_Guido,duan,reply_Bloch,finite_time,QMC_Troyer,QMC_T0}].
To the best of our knowledge, so far all the experimental data
of these non-condensate particles were fitted by a Gaussian type 
distribution function at finite momentum and the obtained fitting 
parameter is interpreted as the temperature of bosons inside 
the optical lattice [\oncite{SF_MI_others}]. This approach may be
justified for weakly interacting bosons at finite temperature,
but it cannot be reliable near the SF-MI transition point, 
where the quantum depletion is known significantly enhanced.
In other words, near the quantum transition point, there will be at least
{\it three} relevant length scales in the interference peak: 
The first one is the condensate size as mentioned above, the 
second one is the thermal wavelength, and the third one is the healing 
length, which is associate with the interaction effect or quantum 
depletion. (The lattice constant
and the size of Wannier function can be assumed not relevant 
when investigating a single peak of the interference pattern.) 
To the best of our knowledge, there is no useful analytical form of 
the momentum distribution to distinguish the two different 
contributions (thermal and quantum) of the non-condensate 
particles near the SF-MI transition point. Exact numerical 
simulation cannot distinguish these two contribution either
[\oncite{QMC_Ohio,QMC_Troyer}]. Without a justified theory to 
describe the momentum distribution of non-condensate particles, 
the measurement and/or determination of the condensate 
fraction in the interference peak also becomes questionable. 
Solving such problem and providing a useful theoretical framework is the 
motivation and the theme of this work.

In this paper, we apply the three-state effective model developed
by Altman {\it et al.} [\oncite{Altman}] to calculate the
finite temperature momentum distribution of quasi-particles near 
the SF-MI transition, where the quantum fluctuation can be comparable
or even stronger than the finite temperature effect.
Our results show that (1) in the superfluid regime, the momentum 
distribution of thermal excited quasi-particles is different from the
one of quantum depleted quasi-particles: the former diverges as 
$|\bfq|^{-2}$ in the small momentum ($\bfq$) limit, while the
latter diverges as $|\bfq|^{-1}$. In other words, 
the noncondensate particle is dominated by thermal 
particles in small momentum regime while it is by quantum 
depletion in the larger momentum regime.
(2) In the weak interacting regime and in the Mott insulator 
regime, we obtained an analytical result
for the momentum distribution at finite temperature, providing the
theoretical fitting equations for the TOF image. The obtained
equations can be used to calculate various physical properties, like 
the condensate fraction, temperature, and single particle gap etc. 
inside the optical lattice. Note that here we just concentrate
on the momentum distribution of particles inside the optical lattice, 
and assume that this distribution can be observed in a long
TOF experiment when the interaction effects during the expansion 
is negligible [\oncite{interaction_TWA}].
For simplicity, we will just consider the uniform system
without the inhomogeneous trapping potential.

This paper is organized as following: In Section \ref{Hamiltonian}
we briefly review the system Hamiltonian and the connection
between momentum distribution inside the optical lattice and the
momentum distribution in free space. The later can be understood as
a result of TOF absorption image if assuming no interaction effect during 
expansion and neglecting the finite size effect [\oncite{finite_time}]. 
In Section
\ref{phase_diagram}, we briefly introduce the three state model
and its meanfield phase diagram obtained near the SF-MI transition
point, both at zero temperature and finite temperature regime. 
In Section \ref{Bogoliubov}, we first study the momentum distribution
calculated in the weakly interacting regime, where one can apply 
Bogoliubov theory to study the general properties of thermal and
quantum depleted quasi-particles. In Section \ref{MI}, we applied
the meanfield solution of the three state model to the Mott insulator
phase and obtain some analytic form of momentum distribution function
at finite temperature regime. Results near the quantum critical
point (QCP) of the SF-MI transition are investigated in Section \ref{QCP}.
In Section \ref{SF}, we further apply the three-state model to
study the momentum distribution in the superfluid side near QCP,
both below and above $T_c$. We then study and compare the momentum 
distribution near SF-MI transition by using the scaling theory of 
quantum $XY$ model in Section \ref{scaling}, and then discuss some
issues related to experimental observation. We finally summarize 
our results in Section \ref{summary}. All the supplementary 
materials about the details of calculation are in the Appendices.

\section{System Hamiltonian and Momentum distribution}
\label{Hamiltonian}

In this paper, we consider bosonic atoms loaded in a 3D square
optical lattice with lattice constant, $a$. 
When the lattice strength is sufficiently large, the system 
can be well described by a single band Hubbard model:
\be
H=-J\sum_{\langle \bfR,\bfR'\rangle}a^\dagger_\bfR a^{}_{\bfR'}
+\frac{U}{2}\sum_\bfR n_\bfR(n_\bfR-1)-\mu\sum_\bfR n_\bfR,
\label{H_HM}
\ee
where $a_\bfR$ is the bosonic field operator at site $\bfR$, and
$n_\bfR=a^\dagger_\bfR a^{}_\bfR$ is the density operator. 
$J$ is the single particle 
tunneling amplitude.
$U$ is onsite interaction energy and $\mu$
is the chemical potential. For simplicity, we neglect the harmonic trapping
potential throughout this paper.
Within this single band approximation, we can express a 
bosonic field operator ($\Psi(\bfr)$) in lattice momentum expansion
(see Appendix \ref{appendix_nk}): 
$\Psi(\bfr)=\sum_\bfk' a_\bfk\Psi_\bfk(\bfr)$, where 
$a_\bfk=\Omega^{-1/2}
\sum_\bfR a_\bfR e^{-i\bfk\cdot\bfR}$ with the lattice volume, 
$\Omega=L^3$ (we assume $L$ lattice sites in each direction).
$\Psi_\bfk(\bfr)$ is the Bloch wavefunction function and $\sum_\bfk'$ 
is the summation of lattice momentum over the first Brilliouin Zone (BZ).
As a result, the free space momentum distribution, $N(\bfq)$,
can be calculated by mapping the field
operator into the plane wave basis (see Appendix \ref{appendix_nk}),
i.e. $N(\bfq)=\frac{1}{\Omega}\sum_{\bfk}'n(\bfk)
\left|\tilde{\Psi}^{}_{\bfk}(\bfq)\right|^2$,
where $\tilde{\Psi}^{}_{\bfk}(\bfq)$ is the Fourier transform of 
Bloch wavefunction $\Psi_\bfk(\bfr)$, and $n(\bfk)=\langle 
a^\dagger_\bfk a^{}_{\bfk}\rangle$ is the momentum distribution 
of the lattice momentum, $\bfk$.

When the lattice potential is sufficiently strong, the lowest band
Wannier function can be well-approximated by a Gaussian function
with width, $\sigma$. 
As a result, the Bloch wavefunction can be easily calculated to be
(see Appendix \ref{appendix_nk}):
$|\tilde{\Psi}_\bfk(\bfq)|^2=f(\bfk-\bfq)|w(\bfq)|^2$, 
where $f(\bfq)\equiv\frac{(2\pi)^3}{\Omega}
\left|\sum_\bfR e^{i\bfq\cdot\bfR}\right|^2=\frac{(2\pi)^3}{\Omega}
\prod_{\alpha}\left|\frac{\sin(L q_\alpha/2)}
{\sin(q_\alpha/2)}\right|^2$
and $|w(\bfq)|^2=\frac{\sigma^3}{\pi^{3/2}}e^{-|\bfq|^2\sigma^2}$. 
The momentum distribution measured in the TOF experiment 
is then just a column integration of $N(\bfq)$ along the $z$ direction, 
leading to (see Appendix \ref{appendix_nk} for details)
$N_\perp(\bfq_\perp)=\int_{-\infty}^\infty dq_z N(\bfq)
\propto\frac{1}{\Omega}\sum_{\bfk_\perp,k_z}{}'n(\bfk)
\left|\tilde{\Psi}^{}_{\bfk_\perp}(\bfq_\perp)\right|^2$
upto an overall constant. Note that to derive above expression, we have
used the fact that $\sigma_z|k_z|\ll 1$ in most regime of the first
BZ. Finally, we note that when the system size is
much larger than the lattice constant, 
$f(\bfq)$ becomes a sharp peaked function
at each reciprocal lattice, $\bfG_\perp$, with width $\sim 2\pi/L$.
As a result, the TOF image, $N_\perp(\bfq_\perp)$,
can be further simplified by approximating $f(\bfq_\perp)$ to be a 
delta function at $G_\perp$, and become (see Appendix \ref{appendix_nk})
\be
N_\perp(\bfq_\perp)&\propto&\frac{1}{L}\sum_{k_z}{}'n(\bfq_\perp,k_z)
|w(\bfq_\perp)|^2
\sum_{\bfG_\perp}\delta(\bfq_\perp-\bfG_\perp)
\nonumber\\
&\propto&n_\perp(\bfq_\perp)|w(\bfq_\perp)|^2\ \ \ \mbox{as }\bfq_\perp\sim 0,
\label{N_perp}
\ee
where $n_\perp(\bfq_\perp)\equiv\frac{1}{L}\sum_{k_z}{}'n(\bfq_\perp,k_z)$
is the momentum distribution integrated over the $z$ direction.

\section{Three-state model and finite temperature phase diagram}
\label{phase_diagram}

\begin{figure}
\includegraphics[width=8cm]{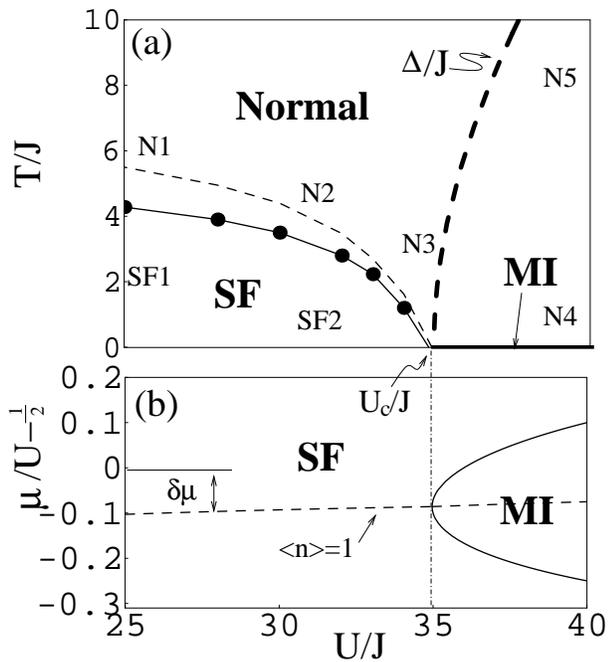}
\caption{
Meanfield phase diagrams of Bose Hubbard model in a 3D square
lattice for $n_0=1$: (a) finite temperature phase diagram and (b) 
quantum phase diagram.
The MI lob in (b) is plotted according to meanfield approximation, with
the lob tip bending down from $U/2$. The
horizontal dashed line in (b) indicates
the particle-hole symmetric line. The finite temperature phase diagram 
in (a) is calculated from Eq. (\ref{total_n_SF2}), and the
critical point at $T=0$ is right the same as the tip position of
the MI lob, as indicated by the vertical dash-dot line. 
Filled dots are the calculated $T_c$ and the thin line is plotted to guide 
the eyes. The thin dashed line above $T_c$ is the line of the constant
peak width, $k_0=0.008\pi/a$, which is the half-width of a Lorenzian fitting
function for the momentum distribution.
The thick dashed lines is the single particle gap energy in the Mott insulator phase. $N1\cdots N5$,
are different regimes of normal state. $SF1$ and $SF2$
are superfluid states in weakly interacting and strongly interacting
regime respectively.
}
\label{phase}
\end{figure}

The phase diagram of the single band Hubbard model has been extensively
studied in the literature in the last ten years. In 2D and 3D systems
with a square lattice, meanfield phase diagram has been shown to be a
very good approximation for the phase boundary [\oncite{QMC_Troyer,strong_coupling,mean_field,finiteT}]. 
The meanfield study of the SF-MI phase transition
can be summarized by the two phase diagrams in Fig. \ref{phase}(a) 
and (b), where the quantum critical point, $U_c$, is defined at 
the tip of the MI lob (with average $n_0$ particle per site) 
in (b). The finite temperature phase
diagram in (a) is obtained by fixing the filling fraction to be an integer,
$\langle n_\bfR \rangle=n_0=1$. 
In order to study the momentum distribution in
the single band Hubbard model, we separate the study into two regimes:
one is the weakly interacting regime of SF phase, and 
the other is the strongly interacting regime, where superfluid order
parameter is very small or even zero when entering the MI regime.
For the convenience of later discussion,
below we briefly outline the underlying meanfield theories for these two
different regimes of phase diagram.

\begin{figure}
\includegraphics[width=6.5cm]{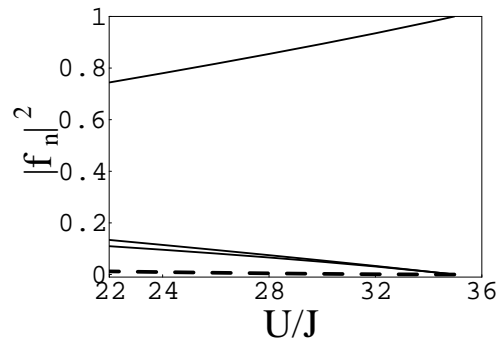}
\caption{
Probabilities of different number occupation, $n$, of each lattice site
in the superfluid state of unit filling ($n_0=\langle a^\dagger_\bfR a^{}_\bfR
\rangle=1$) at $T=0$ (see the text). 
The three solid lines are for $n=1$, $n=0$, and $n=2$
respectively from top to bottom. The dashed line is for $n=3$, and results
for larger number occupation is not visible in this parameter regime.
}
\label{Gutzwiller}
\end{figure}

In the weakly interacting regime, i.e. away from the quantum
phase transition point in the region $N1$ and $SF1$ of 
Fig. \ref{phase}(a), we could apply Bogoliubov-Hartree-Fock-Popov approximation
to study the quasi-particle excitation and quantum depletion effects
both for $T>T_c$ and $T<T_c$. Since this approach has been very well
known in the literature [\oncite{bec_book}], we will just outline
the results and implication to the momentum distribution in the next section.
However, in the second (strongly interacting) regime, we adopt the truncated 
phase space method developed by Altman {\it et al.}
(Ref. [\oncite{Altman}]) to study the superfluid phase near the critical
point and the Mott insulator phase. 
Such truncated phase method is justified
by using the fact that in the strongly interacting regime, number fluctuation
per lattice site can be so small that only three states with number of
particles $n_0$, $n_0+1$, and $n_0-1$, are relevant in the low energy
regime. In order to verify the validity of such truncated phase space
approximation in the superfluid state, we use Gutzwiller trial wavefunction,
$|\Psi_{\rm trial}\rangle=\prod_\bfR \left(\sum_{n=0}^\infty 
f_n|n\rangle_\bfR\right)$, to calculate the probability of higher number
occupation at zero temperature, where $|n\rangle_\bfR$ is the wavefunction
for $n$ particles at site $\bfR$ and $f_n$ is the amplitude of the wavefunction
of $n$ particles at each lattice site. By minimizing the total
energy of Eq. (\ref{H_HM}) with a unit filling fraction ($n_0=1$) in average, 
in Fig. \ref{Gutzwiller}, we find
the probability to have more than two particles per site is very small
($\sim 1\%$, the dashed line) even for $U/J\sim 22$, away from the SF-MI 
transition point, $U_c/J=34.97$. Since the transition temperature is 
also of the same order as the tunneling strength, $J$, it is reasonable 
to expect that such truncated phase space method should be justified in 
a wide range near the SF-MI transition point and at finite temperature 
not too far away from $T_c$.
In this paper, we will name this truncated phase space scheme
to be a three-state model. 
Since details of the theory have been derived in
Ref. [\oncite{Altman}], below we just listed some important results
for completeness and for the convenience of later discussion.
Details of the momentum distribution calculation are shown in the
Appendix \ref{appendix_SF} and \ref{appendix_MI}. 

In the three-state model, one first truncates the phase space to 
$n_0-1$, $n_0$ and $n_0+1$ particles per site (with $n_0$ being the
average filling fraction), by
defining three bosonic operators, $t^\dagger_{\alpha,\bfR}$, to be
$t^\dagger_{\alpha,\bfR}|0\rangle=|n_0+\alpha\rangle_\bfR=
\frac{(a^\dagger_{\bfR})^{n_0+\alpha}}{\sqrt{(n_0+\alpha)!}}
|0\rangle$ for $\alpha=0,\pm 1$. Within this truncated space, a bosonic
creation operator can be expressed to be $a^\dagger_\bfR=
\sqrt{n_0+1}t^\dagger_{1,\bfR}t^{}_{0,\bfR}
+\sqrt{n_0}t^\dagger_{0,\bfR}t^{}_{-1,\bfR}$. One could then rewrite
the original Hamiltonian of Bose-Hubbard model in terms of such new field
operators. The completeness of such truncated phase space is ensured by
applying a constrain: $\sum_{\alpha=-1}^1t^\dagger_{\alpha,\bfR}
t^{}_{\alpha,\bfR}=1$. In order to apply a variational meanfield study
(i.e. Gutzwiller-type ground state), we make a unitary transform
to a coherent state basis: $\vec{t}_\bfR={\cal U}\cdot\vec{b}_\bfR$,
where $\vec{t}_\bfR\equiv [t_{0,\bfR}^\dagger,t^\dagger_{1,\bfR},
t^\dagger_{-1,\bfR}]^T$ and $\vec{b}_\bfR\equiv [b^\dagger_{0,\bfR},
b^\dagger_{1,\bfR},b^\dagger_{2,\bfR}]^T$. 
The transform matrix, ${\cal U}$,
depends on some variational parameters (see Ref. [\oncite{Altman}] and
Appendix \ref{appendix_SF}), 
which are determined by minimizing
the variational energy, $\langle\Psi_{MF}|H|\Psi_{MF}\rangle$,
with $|\Psi_{\rm MF}\rangle\equiv \prod_\bfR b^\dagger_{0,\bfR}|0\rangle$ 
being the trial wavefunction at $T=0$. Such meanfield calculation 
can give a meanfield quantum phase boundary between SF phase and MI phase, as
shown in Fig. \ref{phase}(b). The tip position in (b)
can be found to be at $U_c=zJ(\sqrt{n_0+1}+\sqrt{n_0})^{2}$, 
and $\mu_c=(n_0-\frac{1}{2})U
-\frac{zJ}{2}$ (here $z=6$ is the number of nearest neighboring sites 
in a 3D square lattice). Fixing the average density to be an integer,
$n_0$, the chemical potential at $T=0$ changes as a function of $U$
as the dashed line of Fig. \ref{phase}(b). Defining $\delta\mu=\mu-
(n_0-\frac{1}{2})U$ as the chemical potential deviation from the
middle of MI lob at $U=\infty$, such particle-hole symmetric
(i.e. integer filling) line can be obtained to be:
$\delta\mu=-\frac{1}{4}\left[zJ+U(\sqrt{n_0+1}+\sqrt{n_0})^{-2}\right]$ 
in the SF side and $\delta\mu=-zJ/2$ in the MI side.
The finite temperature phase diagram in (a) is plotted along such 
integer filling line.

To calculate the quasi-particle excitation and the associated quantum
depletion energy, Altman {\it et al.} [\oncite{Altman}]
further used Holstein-Primakoff bosons 
to eliminate $b_{0,\bfR}$ and $b^\dagger_{0,\bfR}$ in order to
obtain an effective Hamiltonian to the quadratic order of 
quasi-particles, $b_{1/2,\bfR}$. In the SF phase, 
the obtained effective Hamiltonian
can be further diagonalized by using a canonical transformation
in momentum space (i.e. a generalized Bogoliubov transformation), 
$\vec{b}_\bfk={\cal M}\cdot\vec{\beta}_\bfk$, 
where $\vec{b}_\bfk\equiv [b^{}_{1,\bfk},b^{}_{2,\bfk},
b^\dagger_{1,\bfk},b^\dagger_{2,\bfk}]^T$ and
$\vec{\beta}_\bfk\equiv [\beta^{}_{s,\bfk},\beta^{}_{m,\bfk},
\beta^\dagger_{s,\bfk},\beta^\dagger_{m,\bfk}]^T$ (see Appendix
\ref{appendix_SF}). Here 
${\cal M}$ is the transformation matrix, and $\beta_{s/m,\bfk}$ are the
quasi-particle field operators for the
sound mode and the massive mode respectively.
In the long wavelength limit, their dispersion becomes
$\epsilon_s(\bfk)=v_s|\bfk|$ and $\epsilon_m(\bfk)
=\Delta_m+\bfk^2/2 m^\ast$, where $v_s$ is sound velocity, $\Delta_m$
is the mass gap and $m^\ast$ is the effective mass of the massive
excitation. Using $\xi=(U_c-U)/J$ to measure the distance 
from the critical point, $U_c$, we find that for integer filling
(along the dashed line of Fig. \ref{phase}(b)):
\be
v_s&=&\frac{\pi\sqrt{n_0(1+n_0)\xi/z}}{(\sqrt{n_0+1}-\sqrt{n_0})^2}
+{\cal O}(\xi^{3/2})
\\
\Delta_m&=&\frac{\sqrt{n_0(1+n_0)}
(1+2n_0)-2n_0(1+n_0)}
{(\sqrt{n_0+1}-\sqrt{n_0})^4)}\xi+{\cal O}(\xi^2)
\nonumber\\
\ee
near the critical point (i.e. $\xi\ll 1$). 
Similar to the ordinary Bogoliubov transformation in the single
component case, the quantum depleted particles can be 
also derived from such canonical
transformation, which is very significant
when near the quantum critical point, $U_c$.
We note that recent calculation [\oncite{spectral_weight}] 
on the spectral weights of these excitations show that the
massive excitation in the superfluid regime has a rather short
life time compared to the low energy phonon excitations. This
result, however, cannot be correctly captured within the 
three-state model used in our recent paper.

In the Mott state, the ground state can be described by the 
three-state model
even better, due to the strong reduction of number fluctuations. 
Within the meanfield approximation, the MI ground state
is still described by
a product state with integer number of particle per site, 
$|\Psi_{MF}\rangle=\prod_\bfR t^\dagger_{0,\bfR}|0\rangle
=\prod_\bfR b^\dagger_{0,\bfR}|0\rangle$. As a result, the obtained
effective Hamiltonian can be easily diagonalized
by another canonical transformation [\oncite{Altman}]:
\be
t_{-1,\bfk}&=&-B(\bfk)\beta^\dagger_{p,\bfk}-A(\bfk)\beta^{}_{h,-\bfk}
\nonumber\\
t_{1,-\bfk}^\dagger &=& A(\bfk)\beta^\dagger_{p,\bfk}
+B(\bfk)\beta^{}_{h,-\bfk},
\label{t-beta}
\ee
where $\beta_{p/h,\bfk}$ is the field operators for 
particle/hole excitation. Here $A(\bfk)\equiv\cosh(D_\bfk/2)$ 
and $B(\bfk)\equiv\sinh(D_\bfk/2)$ are coefficients with 
\be
\tanh(D_\bfk)&\equiv &\frac{-2\epsilon_0(\bfk)\sqrt{n_0(n_0+1)}}
{U-\epsilon_0(\bfk)(2n_0+1)},
\label{D}
\ee
and $\epsilon_0(\bfk)=2J\sum_{\alpha=1}^d\cos(k_\alpha a)$. 
The particle and hole excitation dispersion are given by
\be
\epsilon_{p(h)}(\bfk)=\mp\left[{\epsilon}_0(\bfk)/2+\delta\mu\right]
+\tilde{\omega}(\bfk),
\label{w_ph1}
\ee
where $\delta\mu$ is the chemical potential measured
from $(n_0-\frac{1}{2})U$, ${\epsilon}_0(\bfk)=2J\sum_{\alpha=1}^3
\cos(k_\alpha)$, and 
\be
\tilde{\omega}(\bfk)=
\frac{1}{2}\sqrt{U^2-U{\epsilon}_0(\bfk)(4n_0+2)
+{\epsilon}_0(\bfk)^2}.
\label{w_ph2}
\ee
At zero temperature, the particle and hole
excitations have gaps: 
\be
\Delta\equiv\epsilon_{p}(0)+\epsilon_h(0)
=\sqrt{U^2-zUJ(4n_0+2)+(zJ)^2}.
\label{gap}
\ee
Near the critical point, we have
$\Delta=2(n_0(n_0+1))^{1/4}\sqrt{zJ\delta U}+{\cal O}(U^{3/2})$ 
with $\delta U\equiv U-U_c$. In the strong interacting limit, 
we have $\Delta=U+2z J\sqrt{n_0(n_0+1)}+{\cal O}(U^{-1})$.
In finite temperature, 
the chemical potential ($\delta\mu(T)$) is
given by fixing the number of particles to be $n_0$, i.e. 
number of particles are the same as number of holes: $\frac{1}{\Omega}\sum_\bfk 
\left[f_B(\epsilon_{p,\bfk})-f_B(\epsilon_{h,\bfk})\right]=0$.
More details can be found in Appendix \ref{appendix_MI}.

Using the meanfield ground states and quasi-particle excitations 
described above, in the rest of this paper we will study 
the momentum distribution at finite temperature. Our
results should be able to give a quantitative prediction
for the study of SF-MI quantum phase transition in the experimentally
relevant regime.
 
\section{Momentum distribution in the weakly interacting superfluid phase}
\label{Bogoliubov}

We first study the momentum distribution in the weakly interacting regime
by using the Bogoliubov-Hartree-Fock-Popov approximation [\oncite{bec_book}].
We could separate the calculation in two different regimes of temperature:
(i) $T>T_c$, (2) $T<T_c$. They correspond to 
regions, N1, and SF1 respectively in Fig. \ref{phase}(a).
In the normal state regime, it has been shown [\oncite{bec_book}]
that the self-energy shift to the single particle dispersion 
is just a constant, $\epsilon_\bfk=\epsilon_{b,\bfk}+2n_0U$, where 
$\epsilon_{b,\bfk}=2zJ-\epsilon_0(\bfk)$ is the single particle
band energy measured
from the bottom ($\bfk=0$) and $n_0$ is the average filling fraction. 
As a result, the chemical potential is also shifted by a constant and
the transition temperature, $T_c$, is the same as
the noninteracting system, giving by (set $k_B\equiv 1$)
$n_0=\frac{1}{\Omega}\sum_\bfk'(e^{\epsilon_b(\bfk)/T_c}-1)^{-1}$.
For a 3D square lattice, we have $T_c/J=5.86$ for $n_0=1$
and $T_c/J=10.29$ for $n_0=2$.
When $T$ is very close to $T_c$ from above, the momentum distribution
can be well approximated to be a Lorentzian function:
$n^{\rm N1}(\bfk)\sim\frac{T/J}
{(|\bfk|^2+k_{\rm th}^{\rm N1}{}^2)a^2}$ 
in the long wavelength limit, where $k_{\rm th}^{\rm N1}a
=\sqrt{|\mu(T)-2n_0U|/J}=1.073(\delta T/T_c)^{1/2}$ is the thermal
wavevector and $\delta T\equiv T-T_c$. 
When $T\to T_c$, the thermal momentum $k_{\rm th}^{\rm N1}\ll \pi/a$,
and therefore we could calculate $N_\perp(\bfq_\perp)$ by
integrating $k_z$ from $-\infty$ to $\infty$ as shown in Eq. 
(\ref{N_perp}). We obtain
$N_\perp^{\rm N1}(\bfq_\perp)=\frac{\pi T/t}{\sqrt{|\bfq_\perp|^2
+k_{\rm th}^{\rm N1}{}^2}}|w(\bfq_\perp)|^2$, where 
$w(\bfq_\perp)\propto
e^{-|\bfq_\perp|^2\sigma^2}$ is a broad Gaussian function due to 
the small onsite Wannier function of width $\sigma<a$ within the single
band approximation. As pointed out in Ref. [\oncite{Jason_sharp}],
the TOF image obtained from $N^{\rm N1}_\perp(\bfq_\perp)$ can therefore 
have a ``sharp peak'' even when $T>T_c$.
As a result, such TOF image can be easily mis-interpreted as the 
characteristic feature of condensation. 
This simple analysis shows the importance for an quantitative 
analysis for the TOF image in the optical lattice.

\begin{figure}
\includegraphics[width=8cm]{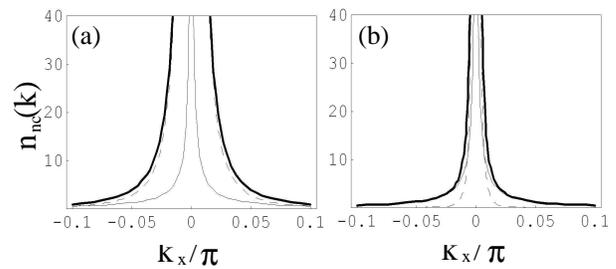}
\caption{
Momentum distribution of non-condensate particles for (a) $T/J=0.2$ and
(b) $T/J=0.02$ in region SF1 of Fig. \ref{phase}(a). 
Thin solid and dashed lines are 
$n^{\rm SF1}_{\rm qn}(\bfk)$ and $n^{\rm SF1}_{\rm th}(\bfk)$
respectively, and thick solid lines are their sum.
Here we use $U/J=0.5$, $n_0=1$ and $k_y=k_z=0$. 
Approximated results of Eq. (\ref{n_SF1}) cannot be distinguished with
the total distribution (thick line).
}
\label{n_k_weak}
\end{figure}

When the temperature is below $T_c$ (i.e. region SF1
of Fig. \ref{phase}(a)), we can treat $a^\dagger_0=a_0$ as
a $c$-number (i.e. condensate at $\bfk=0$) and apply the 
Bogoliubov-Hartree-Fock-Popov approximation 
to diagonalize the quadratic order of the resulting effective 
Hamiltonian. Following the standard approach [\oncite{bec_book}],
we can obtain the phonon dispersion: $\epsilon_B(\bfk)
=\sqrt{\epsilon_b(\bfk)(\epsilon_b(\bfk)+2n_0U_0)}$, and
the momentum distribution can be divided into
two parts: $n^{\rm SF1}(\bfk)=n^{\rm SF1}_{\rm con}\delta(\bfk)
+n^{\rm SF1}_{\rm nc}(\bfk)$, where 
$n^{\rm SF1}_{\rm con}$ is the number density
of condensate particles and $n^{\rm SF1}_{\rm nc}(\bfk)=
n^{\rm SF1}_{\rm th}(\bfk)+n^{\rm SF1}_{\rm qn}(\bfk)$ 
is the momentum distribution of non-condensate particles. Here
$n^{\rm SF1}_{\rm th}(\bfk)=
\frac{\epsilon_b(\bfk)+n_0U_0}{\epsilon_B(\bfk)}f_B(\epsilon_B(\bfk))$
and
$n^{\rm SF1}_{\rm qn}(\bfk)=
\frac{\epsilon_b(\bfk)+n_0U_0}{2\epsilon_B(\bfk)}-\frac{1}{2}$
are the contribution from thermal excitation and quantum 
depletion respectively. $f_B(x)\equiv (e^{x/T}-1)^{-1}$ is 
Bose-Einstein distribution function.
Since the condensate part is a structure-less
delta function in momentum space, here we concentrate on the momentum
distribution of non-condensate particles.
In Fig. \ref{n_k_weak}, we show typical results of 
$n^{\rm SF1}_{\rm th}(\bfk)$ and $n^{\rm SF1}_{\rm qn}(\bfk)$ for comparison.
We note that, even {\it without} adding the Wannier function, 
$n^{\rm SF1}_{\rm th}(\bfk)$
and $n^{\rm SF1}_{\rm qn}(\bfk)$ can have different shape 
in both long and short wavelength limit: 
as shown in Fig. \ref{n_k_weak}(b), $n^{\rm SF1}_{\rm nc}(\bfk)$
is dominated by thermal excitation in the long wavelength limit, while
dominated by quantum depletion in the large momentum regime.
This feature also makes it easy to be mis-interpreted as a ``bimodal
structure'' of condensate, although we do not include the condensate 
particles yet. This situation will become more serious and crucial when
near the quantum critical point, where the number
of condensate particles can be small while the ``bimodal structure''
still exists due to the different momentum distribution 
of thermal and quantum excitation.

To further study the underlying physics in this regime, 
we note that there are two momentum scales involved in above
expression: one is thermal momentum,
$k^{\rm SF1}_{\rm th}a\equiv \sqrt{T/J}$ and the other is quantum momentum,
$k^{\rm SF1}_{\rm qn}a\equiv \sqrt{n_0U/2J}$. 
There are three regimes we need to
identify: (i) In weak interaction limit, we have 
$k^{\rm SF1}_{\rm qn}\ll k^{SF1}_{\rm th}$
and hence the most relevant momentum regime is
$k^{\rm SF1}_{\rm qn}\ll |\bfk|\ll k^{\rm SF1}_{\rm th}$, 
dominated by the thermal
excitation. As a result, using $\epsilon_B(\bfk)\sim\epsilon_b(\bfk)$
for $|\bfk|>k^{\rm SF1}_{\rm qn}$,
we find $n^{\rm SF1}_{\rm nc}(\bfk)\sim n^{\rm SF1a}_{\rm th}(\bfk)\sim
\frac{k^{\rm SF1}_{\rm th}{}^2}{|\bfk|^2}$ in this region. 
(ii) In extremely low temperature region, we have
$k^{\rm SF1}_{\rm th}\ll k^{\rm SF1}_{\rm qn}$, and the 
most relevant momentum regime becomes
$k^{\rm SF1}_{\rm th}\ll |\bfk|\ll k^{\rm SF1}_{\rm qn}$, 
dominated by the quantum depletion
instead. Neglecting the thermal contribution in this region, we find
$n^{\rm SF1}_{\rm nc}(\bfk)\sim n^{\rm SF1}_{\rm qn}(\bfk)\sim 
\frac{k^{\rm SF1}_{\rm qn}{}^2}{4|\bfk|}$
(iii) In the intermediate regime ($U\sim T$),
we have $k^{\rm SF1}_{\rm th}\sim k^{\rm SF1}_{\rm qn}$ and
the most relevant momentum regime is from
$|\bfk|<k^{\rm SF1}_{\rm th},k^{\rm SF1}_{\rm qn}$.
The the leading contribution of non-condensate particles 
can be obtained by expanding in the long wavelength limit:
\be
n^{\rm SF1}_{\rm nc}(\bfk)&\sim&\frac{(k^{\rm SF1}_{\rm th}a)
^2}{2|\bfk|^2a^2}+c_0-c_2|\bfk|^2a^2
\nonumber\\
&\sim&\frac{(k^{\rm SF1}_{\rm th}a)^2}{2|\bfk|^2a^2}+
\frac{c_0}{1+(c_2/c_0)|\bfk|^2 a^2},
\label{n_SF1}
\ee
where 
\be
c_0&=&\frac{(k^{\rm SF1}_{\rm th}a)^2}{24}+\frac{1}{8}
\frac{(k^{\rm SF1}_{\rm th})^2}{(k^{\rm SF1}_{\rm qn})^2}
+\frac{1}{6}\frac{(k^{\rm SF1}_{\rm qn})^2}{(k^{\rm SF1}_{\rm th})^2}
-\frac{1}{2}
\ee
and 
\be
c_2=\frac{1}{32}\frac{(k^{\rm SF1}_{\rm th})^2}
{(k^{\rm SF1}_{\rm qn})^4}
+\frac{1}{90}\frac{(k^{\rm SF1}_{\rm qn})^4}{(k^{\rm SF1}_{\rm th})^6}
-\frac{(k^{\rm SF1}_{\rm th})^2}{480}
-\frac{1}{12 (k^{\rm SF1}_{\rm th}a)^2}.
\nonumber\\
\ee
Our numerical calculation shows that
Eq. (\ref{n_SF1}) is a very good approximation for the full Bogoliubov
result ($< 1\%$) for $T/J>0.1$ and 
$n_0U/J<10$, covering most superfluid regime that current experiment 
can access. After column integration over $k_z$ (we can extend the
integration range to $(-\infty,\infty)$ if only $k^{\rm SF1}_{\rm th/qn}\ll
\pi$), we find
\be
N^{\rm SF1}_\perp(\bfq_\perp)\sim \left(\frac{(k^{\rm SF1}_{\rm th})^2}
{4|\bfq_\perp|}
+\frac{c_0^{2}/2}{\sqrt{c_2(c_0+c_2|\bfq_\perp|^2)}}\right)|w(\bfq_\perp)|^2
\nonumber\\
\label{N_SF1}
\ee
in this regime. Eq. (\ref{N_SF1}) can be used to fit the experimental
data to obtain the temperature and average filling fraction inside the
optical lattice (We note that the overall amplitude is arbitrary and
therefore only two parameters in Eq. (\ref{N_SF1}) 
are needed). In above analysis, we did not discuss the
structure of condensate part, which is supposed to be a sharp peak
at $\bfq=0$ with width $\pi/L$. The details of the condensate shape
may be strongly affected by the interaction effect in the beginning 
of expansion. On the other hand,
our analysis on the non-condensate particles above provides a quantitative
and analytic equation for experimentalists to study the thermal/quantum 
tails for $\bfq_\perp\neq 0$. These parts is much less affected by 
interaction during the expansion and therefore can be a much more reliable
method to extract the temperature inside optical lattice. 

\section{Momentum distribution in the strongly interacting
superfluid phase near SF-MI transition point}
\label{SF}

When the interaction $U$ is large enough, the quantum depletion
of the condensate particles becomes significant, and therefore 
the Bogoliubov approach cannot be justified, especially when
close to the SF-MI transition point. 
More precisely, for the case of unit filling, $n_0=1$,
the Bogoliubov approach used above
is justified when $U\ll J$ (i.e. $k_{\rm qn}^{SF1}a
=\sqrt{n_0U/2J}\ll 1$), while the SF-MI transition occurs at
$U\sim J$. In the later regime, the number fluctuation at each
lattice size is strongly reduced and the phase fluctuation 
becomes enhanced. In this regime, we can apply the
three-state model to study the SF phase as briefly described
in Section \ref{phase_diagram}. To calculate
the momentum distribution in the optical lattice, 
$\langle a^\dagger_\bfk a^{}_\bfk\rangle$, 
we need to apply a series of transformation to diagonalize the
quantum fluctuation on top of the meanfield result of the
three-state model. Details of the calculation is 
shown in Appendix \ref{appendix_SF}. Here we just shown the
final result of the noncondensate part: 
$n^{\rm SF2}_{\rm nc}(\bfk)=n^{\rm SF2}_{\rm th}(\bfk)+
n^{\rm SF2}_{\rm qn}(\bfk)$, where the thermal 
($n^{\rm SF2}_{\rm th}(\bfk)$) and the quantum depleted
($n^{\rm SF2}_{\rm qn}(\bfk)$) parts are respectively (see 
Eq. (\ref{akak_SF})):
\be
n^{\rm SF2}_{\rm th}(\bfk)
&=&(S_{11}(\bfk)^2+S_{33}(\bfk)^2)f(\epsilon_{s,\bfk})
\nonumber\\
&&+(S_{22}(\bfk)^2+S_{44}(\bfk)^2)f(\epsilon_{m,\bfk})
\label{n_SF2_th}
\\
n^{\rm SF2}_{\rm qn}(\bfk)&=&S_{33}(\bfk)^2+S_{44}(\bfk)^2,
\label{n_SF2_qn}
\ee
where
$S_{ij}(\bfk)$ is the matrix element of a $4\times 4$ matrix defined
in Eq. (\ref{akak_SF0}). The TOF image, $N_\perp(\bfq_\perp)$, 
can be also calculated directly from Eq. (\ref{N_perp})
based on above result.
Unfortunately, the analytic expression for the momentum distribution is too 
complicated to be expressed even in the long wavelength limit. Therefore
in this section we will just show the numerical results in various 
parameter regime.

Before showing the numerical result of momentum distribution,
it is instructive to mention that above result can be used to calculate
the interaction effect of the SF transition temperature, $T_c(U)$,
(as shown in Fig. \ref{phase}(a)), by requiring the conservation of
total number of particles:
\be
\frac{1}{\Omega}\sum_\bfk{}'n_{\rm nc}^{\rm SF2}(\bfk)=n_0.
\label{total_n_SF2}
\ee
In other words, when $T<T_c$, the difference between the total number
of particles and the non-condensate particles can be interpreted as
the condensate particles, providing the superfluidity of the system. When 
the interaction is stronger and/or the temperature is larger, the condensate
density becomes smaller so that the non-condensate particles, contributed
both from the thermal excitation and the quantum depletion, becomes
dominant. Although we just consider the quantum fluctuation to the
quadratic order within the three-state model, the obtained transition
temperature (shown in Fig. \ref{phase}(a)) is qualitatively consistent
with the general picture, decreasing dramatically near the SF-MI
transition point at $T=0$. 

When the temperature is above $T_c$, the three-state model is still justified
if only the temperature is much smaller than the on-site interaction $U$
(i.e. small number fluctuation). 
Although the disappearance of the order parameter (or condensation)
makes the meanfield approach of the three-state model not well-justified, 
but quadratic expansion of the elementary excitaion (i.e. $b_{1/2,\bfR}$) 
in the effective Hmailtonian
can be still a good approximation if the number fluctuation is 
still small due to the strong interaction in the low 
temperature reime (i.e. $U\gg T>T_c$).
As a result, we can still calculate the momentum distribution above $T_c$
by tuning the chemical potential away from the value inside the superfluid 
phase, in order to conserve the total number of particles.
In fact, such deviation of the chemical potential above $T_c$ 
automatically leads to the disappearance of the Goldstein mode 
and then renormalizes the effective mass of the quasi-particle excitation.

\begin{figure}
\includegraphics[width=8.3cm]{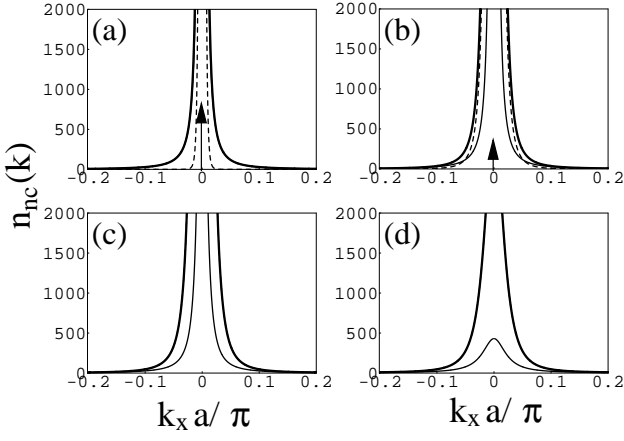}
\caption{
Momentum distribution of non-condensate particles in 3D optical lattice
with $U=34 J$ and $n_0=1$ for different temperature: 
$T/J=0.1$, 1, 1.5, and 1.7 respectively for (a)-(d).
The solid and dashed lines are quantum deplation and thermal particles,
and the thick solid lines are the total. Here $T_c=1.2 J$ 
and the upward arrows in (a) and (b)
indicate the presence of condensate particles at $\bfk=0$.
}
\label{n_k_SF_3state}
\end{figure}

In Fig. \ref{n_k_SF_3state} we show the numerical results 
of the momentum distribution in the optical lattice for the thermal 
and quantum deplated particles in this regime.
Different from the results of weakly interacting regime (SF1), 
the quantum deplation can still dominate the momentum distribution 
of non-condensate particle near the critical point, $U_c$, at low temperature
regime (Fig. \ref{n_k_SF_3state}(a)). When the temperature is approaching
$T_c$ from below, however, the thermal excited particles become dominant
in the momentum distribution (Fig. \ref{n_k_SF_3state}(b)). Similar
to the situation of weakly interacting regime discuss above, momentum
distribution of both quantum deplated and thermal excited particles
are divergent in the long wavelength limit for $T<T_c$, making the
distinguishment between condensate particles and non-condensate particles
highly non-trivial, at least in a uniform system.
When the temperature is above $T_c$, the contribution of condensate particles
disappears and the momentum distribution becomes a Gaussian-like function
with the width increased as a function of temperature.

\section{Momentum distribution in Mott insulator phase}
\label{MI}

Now we move to the study of finite temperature momentum 
distribution in the Mott insulator regime 
(in regions N4, and N5 of Fig. \ref{phase}(a)), 
where the particle number fluctuation at each lattice site 
is strongly reduced. 
As mentioned in Section \ref{phase_diagram}, 
in the three-state model [\oncite{Altman}] the quasi-particle excitation 
can be calculated analytically, leading to a powerful tool to 
study the momentum distribution analytically.
After some straightforward but tedius calculation (details are in Appendix
\ref{appendix_MI}), the full momentum distribution can be also
contributed from thermal part and quantum deplation part as before:
$n^{\rm MI}_{\rm nc}(\bfk)=n^{\rm MI}_{\rm th}(\bfk)+
n^{\rm MI}_{\rm qn}(\bfk)$, where
\be
n^{\rm MI}_{\rm th}(\bfk)
&=&-\delta n+f_B(\epsilon_{p,\bfk})\left(\sqrt{n_0+1}A(\bfk)-\sqrt{n_0}B(\bfk)\right)^2
\nonumber\\
&&+f_B(\epsilon_{h,\bfk})
\left(\sqrt{n_0+1}B(\bfk)-\sqrt{n_0}A(\bfk)\right)^2,
\label{n_MI_th}
\\
n^{\rm MI}_{\rm qn}(\bfk)
&=&\left(\sqrt{n_0+1}B(\bfk)-\sqrt{n_0}A(\bfk)\right)^2
\label{n_MI_qn}
\ee
as shown in Eq. (\ref{nk_MI_lowT}). Here $A(\bfk)$ and $B(\bfk)$
have been defined in Eqs. (\ref{t-beta}) and (\ref{D}) above. 
$\delta n$ a constant density shift due to thermal excitation, and
can be calculated from Eq. (\ref{delta_n}) (see more details in
Appendix \ref{appendix_MI}). 

Since the single particle tunneling, $J$, is much smaller than onsite
interaction, $U$, in this regime, there are only two temperature 
limit we need to consider: 
(i) in the low temperature regime ($U\gg T,J$, i.e. region N4 
of Fig. \ref{phase}(a)), and (ii) in the high teperature regime 
($T\sim U\gg J$, i.e. region N5 of Fig. \ref{phase}(a)).
From the calculation shown in Appendix \ref{appendix_MI},
in the low temperature region ($U\gg T,J$),
the chemical potential can be calculated to be 
(see Eq. (\ref{mu_MI_lowT}):
$e^{\delta\mu/T}=\left(\frac{I_0\left(\frac{2n_0J}{T}\right)}
{I_0\left(\frac{2(n_0+1)J}{T}\right)}\right)^{3/2}$, where $I_0(x)$ is
the modified Bessel function. As a result,
we find the momentum distribution of quantum deplated 
particles and the thermal excited particles can be calculated explicitly
to the leading order of $J/U$ and $T/U$
(see Eq. (\ref{nk_MI_lowT})):
\be
n^{\rm N4}_{\rm qn}(\bfk)&=&n_0+\frac{2n_0(1+n_0)}{U}
{\epsilon}_0(\bfk),
\\
n^{\rm N4}_{\rm th}(\bfk)
&=&g_1e^{(n_0+1)\epsilon_0(\bfk)/T}
+g_2e^{n_0\epsilon_0(\bfk)/T}-\delta n,
\ee
where $g_1\equiv (1+n_0)e^{-U/2T}e^{\delta\mu/T}$ and
$g_2=n_0e^{-U/2T}e^{-\delta\mu/T}$ are small constants 
in this low temperature limit. $\delta n=(2n_0+1)
e^{-U/2T} I_0\left(\frac{2n_0J}{T}\right)^{3/2}
I_0\left(\frac{2(n_0+1)J}{T}\right)^{3/2}$ is a constant
density shift due to thermal excitation as shown in Eq. (\ref{delta_n}).
The TOF image, $N_\perp(\bfq_\perp)$, can be also calculated 
analytically in the leading order of large $U$ to be
(see Eqs. (\ref{N_perp_MI_lowT}) and (\ref{NA_perp_MI_lowT}):
\be
N^{N4}_\perp(\bfq_\perp)&=&\left[n_0+\frac{4n_0(1+n_0)J}{U}
(\cos(q_x)+\cos(q_y))\right.
\nonumber\\
&&-\delta n+G_1 e^{2J(n_0+1)(\cos(q_x)+\cos(q_y))/T}
\nonumber\\
&&\left.+G_2e^{2Jn_0(\cos(q_x)+\cos(q_y))/T}\right]|w(\bfq_\perp)|^2,
\ee
where $G_2=g_1I_0(\frac{2J(n_0+1)}{T})$ and 
$G_2=g_2I_0(\frac{2Jn_0}{T})$.

One interesting thing in the above close form expression is that
the momentum distribution is 
different from the pure meanfield result ($N_\perp(\bfq_\perp)=n_0$) 
by additional contribution from both thermal fluctuation of 
quasi-particles and the quantum deplation. 
The thermal excited particles can contribute a very 
narrow peak when $T\ll J$, 
because $e^{n_0\epsilon_0(q_x,0)/T}\propto e^{2n_0J\cos(q_x a)/T}
\sim e^{2n_0J/T(1-q_x^2a^2/2)}$, showing a characteristic momentum
scale, $\sqrt{T/n_0J a^2}$, in the long wavelength limit.
Although such narrow peak structure
on top of a uniform momentum distribution looks surprising, but it 
is consistent with the sharpe peak structure of a normal state
just above the superfluid $T_c$ in the weakly
interacting regime [\oncite{Jason_sharp}].

\begin{figure}
\includegraphics[width=8cm]{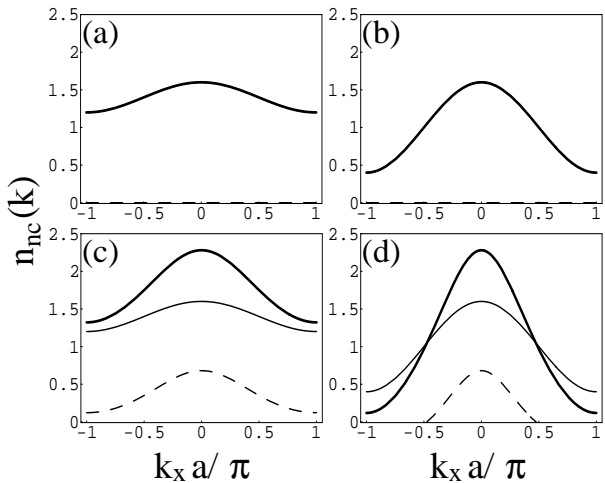}
\caption{
Momentum distribution of non-condensate particles in the MI regime.
(a) and (b) are for $T/J=1$, and (c) and (d) are for $T/J=10$.
The left pannels are disribution along $k_x$ axis with $k_y=k_z=0$, 
while the right pannels are distribution along the diagonal direction,
i.e. $\bfk=k(1,1,1)$. Thin dashed and thin solid lines are 
the thermal and quantum contribution respectively, and the thick 
solid lines are the total distribution.
Here we use $U/J=40$, and other parameters are the same 
as Fig. \ref{n_k_weak}. Note that the thermal contribution in (d) becomes
negative in the large momentum regime due to the constant density shift,
$\delta n$. The total distribution, of course, is always positive definite.
}
\label{n_k_MI_fig}
\end{figure}

Now we consider the high temperature regime when 
$T\sim U\gg J$, as the region N5 in Fig. \ref{phase}.
In this regime, the chemical potential correction,
$\delta\mu$, can be shown to be very small, to the 
second order of $J/T$:
$\frac{\delta\mu}{T}=\frac{-3(2n_0+1)}{2}
\coth\left(\frac{U}{4T}\right)\left(\frac{J}{T}\right)^2$,
as explicitly calculated in Eq. (\ref{mu_high_T}).
This result indicated that the chemical potential, $\mu$,
is almost the same as the value at the middle of MI lob
in the large $U$ limit even in the system of 
low filling fraction ($n_0\sim 1$), reflecting the fact that 
the particle and hole excitations are almost the same
(particle-hole symmetry) in this high temperature limit.
As a result, to the leading order of small single particle
tunneling, $J$, the momentum distribution can be 
calculated to be (see Eq. (\ref{nk_MI_highT})
and Appendix \ref{appendix_MI} for details):
$n^{\rm N5}_{nc}(\bfk)=n^{\rm N5}_{th}(\bfk)
+n^{\rm N5}_{qn}(\bfk)$, where
\be
n^{\rm N5}_{qn}(\bfk)&=&
n_0+2n_0(1+n_0)\frac{\epsilon_0(\bfk)}{U}
\\
n^{\rm N5}_{th}(\bfk)&=&
h_1\frac{\epsilon_0(\bfk)}{U}
+h_2\frac{\epsilon_0(\bfk)}{T}.
\ee
Here we define the two constants, $h_1=\frac{4n_0(1+n_0)}{e^{U/2T}-1}$,
and $h_2=\frac{(1+2n_0+2n_0^2)e^{U/2T}}{(e^{U/2T}-1)^2}$.
Note that the contribution of $\delta n$ is cancelled in the regime,
see Eq. (\ref{nk_MI_highT}).
It is easy to see that in this regime, the thermal excitation
contributes the same momentum distribution ($\propto
\epsilon_0(\bfk)$) as the one contributed by quantum
deplation. This is very different from the result in the
low temperature limit (see $n^{\rm N4}_{\rm th}(\bfk)$
above). This is because, in the low temperature
limit, the thermal wavelength can be very long, leading
to a small momentum distribution of the thermal
excited quasi-particles. However, when the temperature
is increased, the thermal wavelength becomes shorter
and even the same order as the lattice constant, and
hence the momentum distribution is broad in momentum
space. The TOF image obtained by integrating over the $k_z$ 
component can be also calculated easily (see Eqs. 
(\ref{nk_perp_MI_highT}) and (\ref{H_1})):
\be
N_\perp({\bfq_\perp})&=&\left[n_0+H_1 \frac{\epsilon_0(\bfq_\perp)}{J}\right]
|w(\bfq_\perp)|^2
\ee
where $H_1=\frac{2n_0(1+n_0)J}{U}\left[1
+\frac{2}{e^{U/2T}-1}\right]+\frac{(1+2n_0+2n_0^2)e^{U/2T}}{(e^{U/2T}-1)^2}
\frac{J}{T}$.

\section{Momentum distribution at quantum critical point}
\label{QCP}

When near the quatum phase transition point (say N3 of Fig. \ref{phase}), 
the meanfield treatment
becomes unjustified due to large fluctuations. However, we find that
it is still possible to extract some useful information about the momentum
distribution in the large $n_0$ limit, where the meanfield approximation
becomes jusified again, since the transition becomes more classical like.
We can start from the analytic results in MI state by taking 
$U\to U_c$. Using the fact that $A(\bfk)=\cosh(D_\bfk/2)$ and
$B(\bfk)=\sinh(D_\bfk/2)$ with $D_\bfk$ being defined in Eq. (\ref{D}), 
we find that in the large $n_0$ limit:
\be
n_{\rm nc}^{\rm MI}(\bfk)
&=&f_B(\epsilon_{p,\bfk})
\left(\sqrt{n_0+1}A(\bfk)-\sqrt{n_0}B(\bfk)\right)^2-\delta n
\nonumber\\
&&+(1+f_B(\epsilon_{h,\bfk}))
\left(\sqrt{n_0+1}B(\bfk)-\sqrt{n_0}A(\bfk)\right)^2
\nonumber\\
&\sim&n_0(1+f_B(\epsilon_{p,\bfk})+f_B(\epsilon_{h,\bfk}))
(A(\bfk)-B(\bfk))^2.
\nonumber\\
\ee
We also have
\be
&&(A(\bfk)-B(\bfk))^2=e^{-D_\bfk}.
\nonumber\\
&\sim& \sqrt{\frac{U-\epsilon_0(\bfk)(2n_0+1-2\sqrt{n_0(n_0+1)})}
{U-\epsilon_0(\bfk)(2n_0+1+2\sqrt{n_0(n_0+1)})}}
\nonumber\\
&\sim&\sqrt{1-\frac{1}{u}\gamma_\bfk},
\ee
where $u\equiv U/4n_0zJ$ and $\gamma_\bfk\equiv z^{-1}\sum_\delta 
e^{i\bfk\cdot\delta}$ with $\delta$ being the direction to the nearest
neightboring sites [\oncite{Altman}]. 
In other words, for 3D square lattice,
$\epsilon_0(\bfk)\equiv J\gamma_\bfk=
2J\sum_{\alpha}\cos(k_\alpha a)$. Within the same approximation ($n_0\gg 1$),
we have $\epsilon_{p,\bfk}\sim\epsilon_{h,\bfk}=(U/2)
\sqrt{1-\gamma_\bfk/u}$, and therefore
the momentum distribution becomes
\be
n_\bfk^{{\rm MI}} &\propto&(1+2f_B(\epsilon_{p,\bfk}))
\left(1-\frac{1}{u}\gamma_\bfk\right)^{-1/2}
\nonumber\\
&\propto& \coth\left(\frac{\epsilon_{p,\bfk}-\mu}{2k_BT}\right)
\left(1-\frac{1}{u}\gamma_\bfk\right)^{-1/2}
\ee
This is the most general expression of finite temperature
momentum distribution in the MI state in the large $n_0$ limit.

We find that above result can be also easily applied to the
situation at critical point (N3 of Fig. \ref{phase}), where $u=1$ and
$\mu=U/2$ in the large $n_0$ limit.  In the long wavelength limit, we have 
$\gamma_\bfk\sim 1-\frac{1}{6}a^2\bfk^2$, and therefore
\be
n_\bfk^{{\rm N3}}
&\propto&\coth\left(\frac{U}{8k_BT}\right)\frac{1}{|\bfk|a}
\ee
upto some overall density shit, $\delta n$.
When temperature is lowered, the amplitude
becomes larger and hence the sharp peak is more pronounced.
Note that in the low filling factor limit, $n_0\sim 1$, the 
exact cancellation in above derivation will not exist,
and therefore the momentum distribution function will be 
still finite at $\bfk=0$, leading to a very sharp Lorentzian type distribution
with a large correlation length.

\section{Effective theory and the scaling law of the momentum 
distribution near the critical point}
\label{scaling}

The quantum critical theory for the momentum distribution near the
SF-MI quantum critical point at integer filling 
is belong to the univerality class of the $d$+1 dimensional $XY$ model
[\oncite{Fisher}] (here $d$ is spatial dimensionality). 
Before performing a full calculation, several remarks
have to be mentioned: (i) It is known that the superfluid phase 
has only two independent fluctuations and therefore we can
use $XY$ model or quantum rotor model with degree of freedom
$N=2$ to investigate the quantum critical phenomnea near the
SF-MI transition. (ii) Since the {\it anamolous
dimension}, $\eta$, is known to be just 0.03 for $D=2$ dimension
and has a small logarithmic correction for 3D system [\oncite{Sachdev}],
we can expect that this problem can be very well approximated
by the large $N$ result, which is equilvalent to the meanfield 
approximation in the limit of large filling fraction ($n\gg 1$).
(iii) It is well-known that for 2D $XY$ model,
there is a classical Koterlitz-Thouless transition, which has a 
power-law decay of the correlation function in low temperature limit.
The quantum critical phenomena of such quasi-long-ranged order
is highly non-trivial and therefore will not be included in our following
analysis. 

Based on above assumptions, we can consider the quantum field theory
for large $N$ limit and use the value of saddle point to write down
the meanfield order parameters. Following the notation in 
Ref. [\oncite{Sachdev}], the resulting fluctuations of action
above such meanfield state can be written as the following $XY$ model 
(or say the quantum rotor model of $N=2$ degree of freedom):
\be
S&=&\frac{N}{2\tlc\tlg}\sum_\alpha\int d\bfr\int_0^{1/T}d\tau
\left[(\partial_\tau n_\alpha)^2+c^2(\nabla n_\alpha)^2\right.
\nonumber\\
&&\left.+\tlm^2n_\alpha^2\right]
\ee
where $n_{\alpha}$ for $\alpha=1,2$ is the component of 
quantum fluctuations. (The original Hubbard model has $U(1)$ symmetry,
so that a complex order parameter with two indepedent components of quantum
fluctuations are expected.) Here 
$\tlc$ is the renormalized phonon velocity, $\tlg$ is the renormalized
interaction, and $m$ is the renormalized mass term, which changes
sign at the SF-MI phase transition boundary [\oncite{Fisher,Sachdev}]. 
Note that the actual values of
$\tlg$, $\tlc$ and $\tlm$ have to be calculated from a microscopic
theory after integrating over the high energy modes and therefore can be
different from the meanfield results as we derived above within various 
regimes. However, since we are more interested in the universal form of
momentum distribution rather than the quantitative values in this section,
we will just keep them as a fitting parameters that might be able to be
measured from the experimental data.
Besides, in this paper we are interested in the results in the
low energy (or long wavelength) limit,
and therefore we can neglect the lattice potential

Here we first investigate the momentum distribution function
in the normal state, which is a paramagnetic state in the
spin (i.e. quantum rotor model with $N=2$)
language. Using the notation in Ref. [\oncite{Sachdev}], in
the large $N$ theory, the correlation function between different 
components of the quantum rotor is given by the following
fomula:
\be
C_{\alpha\beta}(\bfr,\tau)
&\equiv&\langle n_\alpha(\bfr,\tau)n_\beta(0,0)\rangle
\\
\chi_{\alpha\beta}(k,\omega_n)&\equiv&\int_0^{1/T}d\tau \int d\bfr\,
C_{\alpha\beta}(\bfr,\tau)\,e^{-i\bfk\cdot \bfr+i\omega_n\tau}
\nonumber\\
&=&\chi(k,\omega_n)\delta_{\alpha,\beta}
\label{C_chi}
\\
\chi(k,\omega)&=&
\frac{\tlc\tlg/N}{\tlc^2k^2-(\omega+i\delta)^2+\tlm^2},
\label{chi}
\ee
which is the single particle spectral function, and has been assumed
to be isotropic in the low momentum regime (i.e. when the fluctuation
length scale is much larger than lattice constant). As a result, the 
susceptibility $\chi$ above depends only on the value of momentum, $k=|\bfk|$.
$\omega_n=2\pi nT$ is the Matsubara frequencies, and we have let 
$k_B$ and $\hbar$ to be unit. In the last line of Eq. (\ref{C_chi}), 
we have used the
fact that the correlation function of a paramagnetic state (i.e.
normal state) should have the diagonal part only in the
spin component.

The momentum distribution, $n(k)$, is obtained 
from the equal time correlation and
therefore
\be
n(k)&=&T\sum_{\omega_n}\chi(k,\omega_n)
\nonumber\\
&=&T\sum_{\omega_n}\frac{\tlc\tlg/N}{\tlc^2k^2+\omega_n^2+\tlm^2}.
\ee
Note that different behaviors of $n(k)$ in the different places of
the phase diagram (Fig. \ref{phase}(a)) just related to how
the summation is calculated and how the mass term, $m$, changes in
different regime near the critical point. Using the following
identiy: $\sum_{n=-\infty}^\infty\frac{1}{x^2+n^2}=(\pi/x)
\coth(x\pi)$, we can evaluate the spectral function and obtain:
\be
n(k)&=&\frac{\tlc\tlg/2N}{\sqrt{\tlc^2k^2+\tlm^2}}
\coth\left(\frac{\sqrt{\tlc^2k^2+\tlm^2}}{2T}\right).
\label{n_k_normal}
\ee

When considering large $T$ limit and finite mass term ($\tlm\neq 0$), 
we use $\coth(x)\sim 1/x$ as $x\to 0$, we find
\be
n(k)&=&\frac{\tlc\tlg T/N}{\tlc^2k^2+\tlm^2}=\frac{\tlc\tlg T/\tlm N}
{1+\xi_c^2k^2},
\ee
which is the Lorenzian function one usually discussed in 
high temeprature regime. $\xi_c=\tlc/\tlm$ is the coherent length.
However, if we are looking at low temperature and/or
large momentum regime, Eq. (\ref{n_k_normal}) gives 
\be
n(k)\sim\frac{\tlc\tlg/2N}{\sqrt{\tlc^2k^2+\tlm^2}}\sim k^{-1}
\ee
as $k\gg \tlm/\tlc=\xi_c^{-1}$. This result implies the different momentum
dependence can be obtained when the momentum is larger or smaller
than the inverse of correlation length, which is the length scale
indicating the coherence of the entire system. If $\xi_c$ is large 
(i.e. lower temperature and near the SF transition), most of the momentum distribution 
will be dominated by the quantum fluctuations, $\propto k^{-1}$. 
On the other hand, if $\xi_c$ is small (i.e. away from the SF transition), 
the most momentum distribution will be dominated by thermal 
fluctuations, which gives a Lorentzian type distribution.

Finally, if we consider the momentum distribution along the
the phase transition boundary or inside the ordered phase at finite 
temperature, i.e. $\tlm=0$, we can find that at finite $T$:
\be
n(k)&=&\frac{\tlg/2N}{|k|}\coth\left(\frac{\tlc|k|}{2T}\right)
\nonumber\\
&\sim&\left\{
\begin{array}{lr}
\displaystyle\frac{\tlg T}{Nc\,k^2} & \hspace{1cm}T\gg \tlc k \\
& \\
\displaystyle\frac{\tlg}{2 N |k|} & \hspace{1cm}T\ll ck
\end{array}\right.
\ee
This shows two different behavior for large $T$ (or small $k$)
and small $T$ (or large $k$ regime). This result agree with our
statement from the three state model that thermal fluctuation
is dominant and diverges as $k^{-2}$ in the small momentum 
regime, while the quantum fluctuation is dominant and scales
as $|k|^{-1}$ in the large momentum regime.

\section{Experimental Implication}
\label{discussion}

The TOF image is known to be the most important measurement
in the systems of ultracold atoms. The long time of flight image is
usually believed to be equilivant to the momentum distribution of
the system inside the optical lattice. Our results show that how the
different scaling behavior and temperature/interaction dependence
of such distribution of non-condensate particle can be distinguished
and identified. This is of particular importance when near the quantum
critical point of the SF-MI transition. From experimental point of 
view, there are several methods to observe these different behaviors
in the current experimental setup. 

First of all, one can systematically
investigate the TOF image for different initial temperature, and compare
their TOF image both in the small and large momentum regime. According
to our results, one should find that the TOF in the large momentum 
regime has little temperature dependence since it is mostly contributed
by quantum fluctuations. On the other hand, after extract out the
temperature dependent part of the TOF, one can use our analytical
form of the thermal excitation (Eqs. (\ref{n_SF1}) and (\ref{N_SF1})) to obtain the temperature and condensate density inside the optical lattice. 

In recent experiments in Stanford's group [\oncite{Stanford}], 
the TOF image of a condensate confined in the 1D optical lattice has been
fitted and investigated by using three Gaussian-type function: one for
condensate, one for thermal atoms, and one for quantum depletion. Although 
the system setup and parameter regimes in their experiment are different from
the regimes we discuss in this paper, the clear evidence for different
temperature behaviors in the non-condensate particles still qualitatively
supports our prediction: the quantum depleted particles dominates the large
momentum distribution, while the thermal particles dominates the small
momentum regime. Further investigation on the TOF image can be very interesting
and useful for the future development.

Secondly, one can also separate the contribution of the thermal excitations
from the quantum deplation by providing a potential gradient just before 
releasing the optical lattice. It is found [\oncite{private_ian}] 
that due to the different linear response of these two non-condensate 
particles, the center of thermal particle distribution can be away from the
quantum deplated particles, making the direct measurement of these
two distribution functions available. This approach opens the 
possibility to quantitatively study the condensate and non-condensate
profile near the quantum critical point. 

\section{Summary}
\label{summary}

In this paper, we have explicitly shown that the momentum distribution
of the quantum deplated non-condensate particles can be very
different from the thermal excited noncondensate particles, especially
when near the quantum critical point of SF-MI phase transition.
Our results is also consistent with the general scaling properties derived
from $N=2$ quantum rotor model (i.e. $XY$ model). Our analytical results
can be used to provide a unique method to determine the temperature
and condensate fraction inside the optical lattice. Futher extension to 
the study of noise correlation function and/or other physical properties
can be also expected in the same frame work.

\section{Acknowledgement}

The author appreciate the fruitful discussion with T.-L. Ho,
T. Porto, I. Spielman, S. D. Huber, S. Sachedev 
and E. Demler. This work was initiated by the discussion with T.-L. Ho
during the visit in Ohio State University. This work is supported by NSC Taiwan.

\appendix

\section{Momentum distribution in an optical lattice and in free space}
\label{appendix_nk}

\subsection{Definition of wavefunction and their Fourier transform}

We first identify the following four kinds of single particle wavefunctions
in the optical lattice and clarify their relationship with each other.
(i) $\Psi_\bfk(\bfr)$ is the Bloch wavefunction in real space at a 
lattice quantum number, $\bfk$, in the first Brillouin Zone (BZ). 
(ii) $\tilde{\Psi}_\bfk(\bfq)$ is its Fourier transform with respect
to the real space coordinate, $\bfr$. (iii) $w_\bfR(\bfr)$ is
the Wannier function, centered at a given lattice position, $\bfR$.
$w_\bfR(\bfr)$ is a Fourier transform of $\Psi_\bfk(\bfr)$ with respect
to the lattice momentum, $\bfk$. (iv) Finally, $\tilde{w}_\bfR(\bfq)$ 
is the Fourier transform of $w_\bfR(\bfr)$ with respect to the real 
space coordinate, $\bfr$.
Their relationship are given as following ($L$ is the number of total lattice
point along each direction and $a$ is
the lattice constant. $\sum_\bfk{}'$ means a summation over the 
first BZ):
\be
\tilde{\Psi}_\bfk(\bfq)&=&\int d\bfr\,e^{-i\bfq\cdot\bfr}
\Psi_\bfk(\bfr)
\label{tilde_Psi}
\\
\Psi_\bfk(\bfr)&=&\int\frac{d\bfq}{(2\pi)^3} e^{i\bfq\cdot\bfr}
\tilde{\Psi}_\bfk(\bfq)
\\
\nonumber\\
w_\bfR(\bfr)&=&\frac{1}{L^{3/2}}\sum_\bfk{}' e^{-i\bfk\cdot\bfR}\Psi_\bfk(\bfr)
\\
\Psi_\bfk(\bfr)&=&\frac{1}{L^{3/2}}\sum_\bfR e^{i\bfk\cdot\bfR}w_\bfR(\bfr)
\\
\nonumber\\
\tilde{w}_\bfR(\bfq)&=&\int d\bfr e^{-i\bfq\cdot\bfr}
w_\bfR(\bfr)
\\
w_\bfR(\bfr)&=&\int\frac{d\bfq}{(2\pi)^3} e^{i\bfq\cdot\bfr}
\tilde{w}_\bfR(\bfq)
\\
\nonumber\\
\tilde{w}_\bfR(\bfq)&=&\frac{1}{L^{3/2}}\sum_\bfk{}' e^{-i\bfk\cdot\bfR}
\tilde{\Psi}_\bfk(\bfq)
\\
\tilde{\Psi}_\bfk(\bfq)&=&\frac{1}{L^{3/2}}\sum_\bfR e^{i\bfk\cdot\bfR}
\tilde{w}_\bfR(\bfq)
\ee
Note that the Fourier and the inverse Fourier transform convention is different
for real space momentum and lattice momentum.
The amplitude of a wavefunction can be shown to be normalized:
\be
1&=&\int d\bfr |\Psi_\bfk(\bfr)|^2=\int\frac{d\bfq}{(2\pi)^3} 
|\tilde{\Psi}_\bfk(\bfq)|^2
\nonumber\\
&=&\int d\bfr |w_\bfR(\bfr)|^2=\int\frac{d\bfq}{(2\pi)^3}
|\tilde{w}_\bfR(\bfq)|^2
\ee
%

\subsection{Momentum distribution}

Now we need to identify the field operators based on the following 
three different basis:
(i) lattice site basis ($\bfR$) with Wannier function ($w_\bfR(\bfr)$)
as an eigenstate, 
(ii) Bloch momentum basis ($\bfk$) with Bloch wavefunction ($\Psi_\bfk(\bfr)$)
as an eigenstate, and
(3) free space momentum basis ($\bfq$) with plane waves ($\frac{e^{i\bfq\cdot\bfr}}{\sqrt{V}}$, and $V$ is the volume of 
the whole system of free space) as an eigenstate. Their fields operators
are defined to be $a_\bfR$, $a_\bfk$ and $b_\bfq$, respectively, and they
are related to the single particle field operator, $\hat{\Psi}(\bfr)$, to be
\be
\hat{\Psi}(\bfr)&=&\sum_\bfR a_\bfR w_\bfR(\bfr)
\nonumber\\
&=&\sum_\bfk{}'a_{\bfk} \Psi_\bfk(\bfr)
\nonumber\\
&=&\sum_\bfq b_\bfq 
\,\frac{e^{i\bfq\cdot\bfr}}{\sqrt{V}}.
\ee
Therefore, these field operators can be transformed with each other 
as following:
\be
a_{\bfk}&\equiv&\frac{1}{L^{3/2}}\sum_\bfR a_\bfR
e^{-i\bfk\cdot\bfR}
\\
b_\bfq&=&\frac{1}{\sqrt{V}}\sum_\bfk 
a_\bfk \tilde{\Psi}_\bfk(\bfq).
\ee

As a result, the momentum distribution in real space ($N(\bfq)$)
can be connected to the momentum distribution in lattice 
($n(\bfk)$) by
\be
N(\bfq)&\equiv&\langle b^\dagger_\bfq b^{}_\bfq\rangle
\nonumber\\
&=&\frac{1}{V}\sum_{\bfk}'n(\bfk)
\left|\tilde{\Psi}^{}_{\bfk}(\bfq)\right|^2,
\label{N_q}
\ee
where we have used $\langle a^\dagger_\bfk a^{}_{\bfk'}\rangle
=n(\bfk)\delta_{\bfk,\bfk'}$.

It is known that when the lattice strength is strong, we can approximate
the Wannier function of the lowest subband by a Gaussian-type function:
\be
w_\bfR(\bfr)&\approx&\frac{1}{\pi^{3/4}\sigma^{3/2}}
e^{-\frac{1}{2}|\bfr-\bfR|^2/\sigma^2}
\ee
where $\sigma$ is the Gaussian width. 
Therefore, we have
\be
|\tilde{\Psi}_\bfk(\bfq)|^2&=&
\left|\frac{1}{L^{3/2}}\sum_\bfR\tilde{w}_\bfR(\bfq)\,e^{i\bfk\cdot\bfR}
\right|^2
\nonumber\\
&=&f(\bfk-\bfq)|w(\bfq)|^2,
\label{Psi_k_q}
\ee
where 
\be
f(\bfq)&\equiv&\frac{(2\pi)^3}{L^3}\left|\sum_\bfR e^{i\bfq\cdot\bfR}\right|^2
\nonumber\\
&=&\frac{(2\pi)^3}{L^3}
\prod_{\alpha}\left|\frac{\sin(Lq_\alpha a/2)}{\sin(q_ja/2)}\right|^2
\label{f_q}
\\
|w(\bfq)|^2&\equiv&\frac{\sigma^3}{\pi^{3/2}}e^{-|\bfq |^2\sigma^2}
\label{w_q}
\ee
In the limit of large system size ($L\to\infty$),  $f(\bfq)$ becomes a
periodic function of delta-fuction peaks at each reciprocal lattice,
$\bfG_\bfn=(2\pi\bfn/a)$, where $\bfn=(n_1,n_2,n_3)$ is a vector
of integer numbers. Within
such approximation, the momentum distribution in free space can have a 
simple relationship with the momentum distribution in the optical
lattice:
\be
N(\bfq)&\propto&\sum_\bfn\sum_\bfk{}'\left[n(\bfk) |w(\bfq)|^2
\delta(\bfk-\bfq-\bfG_\bfn)\right].
\ee
However, such simplified result is not correct if the system of 
finite size, where $f(\bfq)$ is not a delta function, but just a
sharp function at $\bfG_\bfn$.

\subsection{Momentum distribution after column integration}

The TOF image taking after a long expansion time can be understood
as the momentum distribution after column integration along a certain
direction (say $z$ axis). Therefore, it is instructive to study the 
momentum distribution in free space after column integration along
the $z$ axis. From a general result of Eqs. (\ref{Psi_k_q})-(\ref{w_q}), 
we have
\be
N_\perp(\bfq_\perp)&=&\frac{1}{V}
\sum_\bfk{}'n(\bfk)\int\frac{dq_z}{2\pi}
\left|\tilde{\Psi}_\bfk(\bfq)\right|^2
\nonumber\\
&=&\frac{1}{V}\sum_\bfk{}'n(\bfk)
\frac{(2\pi)^3}{L^3}\frac{\sigma^3}{\pi^{3/2}}
\nonumber\\
&&\times\int\frac{dq_z}{2\pi}
\sum_{\bfR,\bfR'}e^{i(\bfk-\bfq)\cdot(\bfR-\bfR')
-\bfq^2\sigma^2}
\nonumber\\
&\propto&\frac{1}{V}\sum_\bfk{}'n(\bfk)
\left|\tilde{\Psi}_{\bfk_\perp}(\bfq_\perp)\right|^2
\nonumber\\
&&\times
\sum_{R_z,R_z'}e^{ik_z(R_z-R'_z)}e^{-(R_z-R'_z)^2/4\sigma^2},
\ee
where $\tilde{\Psi}_{\bfk_\perp}(\bfq_\perp)$ is the Fourier transform of
the 2D Bloch function in 2D lattice, similar to the 3D case
defined in Eqs. (\ref{tilde_Psi}), (\ref{Psi_k_q}), (\ref{f_q}), and (\ref{w_q}).

To evaluate the summation over $R_z$ and $R_z'$, we note that the
first order cotribution comes from the case $R_z=R_z'$, while the second
order contribution comes from $|R_z-R_z'|=a$, which is  exponentially
smaller the first order as $a\gg \sigma$ near the SF-MI transition regime.
As a result, it will be a good approximation to keep the leading order
contribution only and hence the momentum distribution in free space
after column integration becomes
\be
N_\perp(\bfq_\perp)&\propto &
\sum_{\bfk_\perp}{}' n_{\perp}({\bfk_\perp})
\left|\tilde{\Psi}_{\bfk_\perp}(\bfq_\perp)\right|^2
\nonumber\\
&\propto &
n_{\perp}({\bfk_\perp})|w(\bfq_\perp)|^2,
\ee
if we are interested in the long wavelength limit ($\bfq\sim 0$)
and approximate the function $f(\bfq_\perp)\propto\delta(\bfq_\perp-\bfk_\perp)$
in the limit of infinite number of lattice ($L\to\infty$). Here
\be
n_{\perp}({\bfk})\equiv\frac{1}{L}\sum_{k_z}{}' n(\bfk)
\label{n_perp}
\ee
is the momentum distribution in optical lattice after column
integration.

\section{Finite temperature momentum distribution in superfluid 
phase}
\label{appendix_SF}

\subsection{Canonical transformation}

In order to clarify the complicated process, here we list 
all the transformation of quasi-particles used in the SF 
phase within the three-state model. 
Within the three-state model, Altman {\it et al.} 
[\oncite{Altman}] introduced a new operator,
$t_{\alpha,\bfR}^\dagger |0\rangle\equiv 
\frac{1}{\sqrt{(n_0+\alpha)!}}
\left(a^\dagger_\bfR\right)^{n_0+\alpha}|0\rangle=
|n_0+\alpha\rangle_\bfR$
to create a state of $n_0+\alpha$ particles 
($\alpha=\pm 1$ and 0) at site $\bfR$. As a result, one can show that
within this trucated space, any local operator, say
the bosonic field operator $a_\bfR$, and the particle number operator, 
$a^\dagger_\bfR a^{}_\bfR$, can be respectively expressed to be
\be
a_\bfR&=&\sqrt{n_0+1}t^\dagger_{0,\bfR}t^{}_{1,\bfR}
+\sqrt{n_0}t^\dagger_{-1,\bfR}t^{}_{0,\bfR}
\\
a^\dagger_\bfR a^{}_\bfR &=& (n_0+1)t^\dagger_{1,\bfR}t^{}_{1,\bfR}
+n_0t^\dagger_{0,\bfR}t^{}_{0,\bfR}
\nonumber\\
&&+(n_0-1)t^\dagger_{-1,\bfR}t^{}_{-1,\bfR}.
\label{aRaR}
\ee
We note that in the three-state presentation, 
the number operator, $a^\dagger_\bfR a^{}_\bfR$, is not a simple
product of the result for $a_\bfR$ and $a^\dagger_\bfR$.
This can be easily understood from the fact that when
applying $a^\dagger_\bfR a^{}_\bfR$ on a state 
$|n_0-1\rangle_\bfR$, the particle number 
can be virtually reduced to $n_0-2$ by $a^{}_\bfR$ first then becomes
its original value by $a^\dagger_\bfR$. such virtual process is not 
included in the three-state representation of $a_\bfR$.
Such trivial notational problem occurs only in the situation when
two or more field operators applying on the same lattice site.

Within the meanfield approximation, we can assume that
the SF ground state is a coherent state composed by different 
number of states at each site, i.e.
$|\Psi_{\rm MF}\rangle\equiv\prod_\bfR b^\dagger_{0,\bfR}|0\rangle$,
where $b_{0,\bfR}^\dagger=\cos(\theta/2)t^\dagger_{0,\bfR}
+\sin(\theta/2)(\cos\chi t^\dagger_{1,\bfR}+\sin\chi 
t^\dagger_{-1,\bfR})$ where $\theta$ and $\chi$ are two variational
parameters [\oncite{Altman}]. When considering the quasi-particle
excitation, Altman {\it et al.} further generalize above transformation
to the following:
\wbe
\be
\left[\begin{array}{c}
t^\dagger_{0,\bfR} \\ t^\dagger_{1,\bfR} \\ t^\dagger_{-1,\bfR}
\end{array}\right]
&=&\left[\begin{array}{ccc}
\cos(\theta/2) & -\sin(\theta/2)& 0 \\
\sin(\theta/2)\cos\chi & \cos(\theta/2)\cos\chi & -\sin\chi \\
\sin(\theta/2)\sin\chi & \cos(\theta/2)\sin\chi & \cos\chi
\end{array}\right]
\left[\begin{array}{c}
b^\dagger_{0,\bfR} \\ b^\dagger_{1,\bfR} \\ b^\dagger_{2,\bfR}
\end{array}\right]
\nonumber\\
&\equiv & \left[\begin{array}{ccc}
U_{00} & U_{01} & U_{02} \\
U_{10} & U_{11} & U_{12} \\
U_{20} & U_{21} & U_{22} \\
\end{array}\right]
\left[\begin{array}{c}
b^\dagger_{0,\bfR} \\ b^\dagger_{1,\bfR} \\ b^\dagger_{2,\bfR}
\end{array}\right]
\equiv {\cal U}\cdot \vec{b}_\bfR,
\label{cal_U_def}
\ee
\wee
where ${\cal U}$ is the transform matrix, and $\vec{b}_\bfR
\equiv [b^\dagger_{0,\bfR},b^\dagger_{1,\bfR},b^\dagger_{2,\bfR}]^T$
is the vector of operators with $b_{1/2,\bfR}$ being the field 
operators for the excitations above the condensate state, 
$b_{0,\bfR}^\dagger$. 

When in the SF state, one can approximate the single particle ground
state as a $c$-number, and apply the generalized bogoliubov tranformation
to diagonalize the system Hamiltonian upto the qudratic order of 
excitations. Such generalized Bogoliubov transformation can be written
to be [\oncite{Altman}]
\wbe
\be
\vec{b}_\bfk\equiv\left[\begin{array}{c}
b^{}_{1,\bfk} \\ b^{}_{2,\bfk} \\ b^\dagger_{1,-\bfk} \\ b^\dagger_{2,-\bfk}
\end{array}\right]
&=&\left[\begin{array}{cccc}
N_{11}(\bfk) & N_{12}(\bfk) & P_{11}(\bfk) & P_{12}(\bfk) \\
N_{12}(\bfk) & N_{22}(\bfk) & P_{12}(\bfk) & P_{22}(\bfk) \\
P_{11}(-\bfk) & P_{12}(-\bfk) & N_{11}(-\bfk) & N_{12}(-\bfk) \\
P_{12}(-\bfk) & P_{22}(-\bfk) & N_{12}(-\bfk) & N_{22}(-\bfk)
\end{array}\right]
\left[\begin{array}{c}
\beta^{}_{s,\bfk} \\ \beta^{}_{m,\bfk} \\ \beta^\dagger_{s,-\bfk} 
\\ \beta^\dagger_{m,-\bfk}
\end{array}\right]
\equiv {\cal M}\cdot \vec{\beta}_\bfk
\label{b-beta}
\ee
\wee
where the definition of the excited states, $\vec{b}_\bfk$ and 
$\vec{\beta}_\bfk$ are clearly shown as above. ${\cal M}(\bfk)$ is
the $4\times 4$ transform matrix, which in principle can be derived
analytically. However, since the full analytic expression is too complicated
to be understood in a simple physical picture, here we will use numerical 
method to evaluate them directly.

\subsection{Momentum distribution of noncondensate particles}

Our goal is to calculate the momentum distribution in optical lattice:
$\langle a_\bfk^\dagger a_{\bfk}\rangle$, by usng above transformations.
We first transform it into the $t$-operators:
\wbe
\be
\langle a_\bfk^\dagger a_{\bfk}\rangle 
&=&\frac{1}{L^3}\sum_{\bfR,\bfR'}e^{i\bfk\cdot(\bfR-\bfR')}
\left[(n_0+1)\langle t^\dagger_{1,\bfR}t^{}_{0,\bfR}
t^\dagger_{0,\bfR'}t^{}_{1,\bfR'}\rangle
+n_0\langle  t^\dagger_{0,\bfR}t^{}_{-1,\bfR} 
t^\dagger_{-1,\bfR'}t^{}_{0,\bfR'}\rangle\right.
\nonumber\\
&&\left.+\sqrt{n_0(n_0+1)}\left(
\langle t^\dagger_{1,\bfR}t^{}_{0,\bfR}
t^\dagger_{-1,\bfR'}t^{}_{0,\bfR'}\rangle
+\langle t^\dagger_{0,\bfR}t^{}_{-1,\bfR} 
t^\dagger_{0,\bfR'}t^{}_{1,\bfR'}\rangle\right)\right]
\label{akak}
\ee
The first term, defined to be $T_1$ upto the prefactor, $n_0+1$, gives
\be
&&\frac{1}{L^3}\sum_{\bfR,\bfR'}e^{i\bfk\cdot(\bfR-\bfR')}
\langle t^\dagger_{1,\bfR}t^{}_{0,\bfR}t^\dagger_{0,\bfR'}t^{}_{1,\bfR'}\rangle
=L^3\delta_{\bfk,0}U_{10}^2U_{00}^2
\nonumber\\
&&+U_{10}U_{00}\delta_{\bfk,0}\sum_\bfR\left[
\langle(U_{01}b^\dagger_{1,\bfR}+U_{02}b^\dagger_{2,\bfR})
(U_{11}b^{}_{1,\bfR}+U_{12}b^{}_{2,\bfR})\rangle
+\langle (U_{11}b^\dagger_{1,\bfR}+U_{12}b^\dagger_{2,\bfR})
(U_{01}b^{}_{1,\bfR}+U_{02}b^{}_{2,\bfR}) \rangle\right]
\nonumber\\
&&+\frac{1}{L^3}\sum_{\bfR,\bfR'}e^{i\bfk\cdot(\bfR-\bfR')}
\left[U_{10}^2\langle (U_{01}b^{}_{1,\bfR}+U_{02}b^{}_{2,\bfR})
(U_{01}b^\dagger_{1,\bfR'}+U_{02}b^\dagger_{2,\bfR'})\rangle
+U_{00}^2\langle (U_{11}b^\dagger_{1,\bfR}+U_{12}b^\dagger_{2,\bfR})
(U_{11}b^{}_{1,\bfR'}+U_{12}b^{}_{2,\bfR'})\rangle\right]
\nonumber\\
&&+\frac{U_{10}U_{00}}{L^3}
\sum_{\bfR,\bfR'}e^{i\bfk\cdot(\bfR-\bfR')}\left[\langle 
(U_{01}b^{}_{1,\bfR}+U_{02}b^{}_{2,\bfR})
(U_{11}b^{}_{1,\bfR'}+U_{12}b^{}_{2,\bfR'})\rangle
+\langle (U_{11}b^\dagger_{1,\bfR}+U_{12}b^\dagger_{2,\bfR})
(U_{01}b^{\dagger}_{1,\bfR'}+U_{02}b^{\dagger}_{2,\bfR'})
\rangle\right],
\nonumber\\
\ee
\wee
where we have used the approximation:
$b_{0,\bfR}\sim 1-\frac{1}{2}b^\dagger_{1,\bfR}b^{}_{1,\bfR}
-\frac{1}{2}b^\dagger_{2,\bfR}b^{}_{2,\bfR}$ to eliminate $b_{0,\bfR}$
to the quadratic order of small fluctuation, $b_{1/2,\bfR}$.
The first two terms are proportional to the system size, $L^3$, 
and are nonzero only at $\bfk=0$, indicating a contribution from 
the condensate particles. The last two terms, however,
are the noncondensate part. Since in our paper, we are interested
in the momentum distribution at finite lattice momentum $\bfk$, 
we will concentrate on the later part for the numerical 
calculation of momentum distribution.

Below we define the momentum distribution of the noncondensate 
particle from the $i$th term of  Eq. (\ref{akak}) to be $T_i$ ($i=1,2,3,4$)
(not including the prefactor, $n_0$, $n_0+1$, or $\sqrt{n_0(n_0+1)}$),
and these terms can be expressed in the momentum space to be
\wbe
\be
T_1&=&\left\langle\left[\begin{array}{c}
b^{\dagger}_{1,\bfk} \\ b^{\dagger}_{2,\bfk} 
\\ b^{}_{1,-\bfk} \\ b^{}_{2,-\bfk}\end{array}\right]^T
\left[\begin{array}{cccc}
U_{00}^2U_{11}^2 & U_{00}^2 U_{11}U_{12} & U_{10}U_{00}U_{11}U_{01} &
U_{10}U_{00}U_{11}U_{02} \\
U_{00}^2 U_{11}U_{12} & U_{00}^2U_{12}^2 & U_{10}U_{00}U_{12}U_{01} &
U_{10}U_{00}U_{12}U_{02} \\
U_{10}U_{00}U_{01}U_{11} & U_{10}U_{00}U_{01}U_{12} & U_{10}^2U_{01}^2 &
U_{10}^2U_{01}U_{02} \\
U_{10}U_{00}U_{02}U_{11} & U_{10}U_{00}U_{02}U_{12} & U_{10}^2U_{01}U_{02} &
U_{10}^2U_{02}^2
\end{array}\right]
\left[\begin{array}{c}
b^{}_{1,\bfk} \\ b^{}_{2,\bfk} \\ b^\dagger_{1,-\bfk} \\ b^\dagger_{2,-\bfk}
\end{array}\right]\right\rangle
\\
T_2&=&\left\langle\left[\begin{array}{c}
b^{\dagger}_{1,\bfk} \\ b^{\dagger}_{2,\bfk} 
\\ b^{}_{1,-\bfk} \\ b^{}_{2,-\bfk}\end{array}\right]^T
\left[\begin{array}{cccc}
U_{20}^2U_{01}^2 & U_{20}^2 U_{02}U_{01} & U_{20}U_{00}U_{01}U_{21} &
U_{20}U_{00}U_{01}U_{22} \\
U_{20}^2 U_{02}U_{01} & U_{20}^2U_{02}^2 & U_{20}U_{00}U_{02}U_{21} &
U_{20}U_{00}U_{02}U_{22} \\
U_{20}U_{00}U_{21}U_{01} & U_{20}U_{00}U_{21}U_{02} & U_{10}^2U_{21}^2 &
U_{00}^2U_{21}U_{22} \\
U_{20}U_{00}U_{22}U_{01} & U_{20}U_{00}U_{22}U_{02} & U_{00}^2U_{21}U_{22} &
U_{00}^2U_{22}^2
\end{array}\right]
\left[\begin{array}{c}
b^{}_{1,\bfk} \\ b^{}_{2,\bfk} \\ b^\dagger_{1,-\bfk} \\ b^\dagger_{2,-\bfk}
\end{array}\right]\right\rangle
\\
T_3&=&\left\langle\left[\begin{array}{c}
b^{\dagger}_{1,\bfk} \\ b^{\dagger}_{2,\bfk} 
\\ b^{}_{1,-\bfk} \\ b^{}_{2,-\bfk}\end{array}\right]^T
\left[\begin{array}{cccc}
U_{00}U_{20}U_{11}U_{01} & U_{00}U_{20}U_{11}U_{02} 
& U_{00}^2U_{11}U_{21} & U_{00}^2U_{11}U_{22} \\
U_{00}U_{20}U_{12}U_{01} & U_{00}U_{20}U_{12}U_{02} 
& U_{00}^2U_{12}U_{21} & U_{00}^2U_{12}U_{22} \\
U_{10}U_{20}U_{01}^2 & U_{10}U_{20}U_{02}U_{01} 
& U_{10}U_{00}U_{01}U_{21} & U_{10}U_{00}U_{01}U_{22} \\
U_{10}U_{20}U_{01}U_{02} & U_{10}U_{20}U_{02}^2 
& U_{10}U_{00}U_{02}U_{21} & U_{10}U_{00}U_{02}U_{22}
\end{array}\right]
\left[\begin{array}{c}
b^{}_{1,\bfk} \\ b^{}_{2,\bfk} \\ b^\dagger_{1,-\bfk} \\ b^\dagger_{2,-\bfk}
\end{array}\right]\right\rangle,
\\
\ee
\wee
and the forth term ($T_4$) is the exactly the complex conjugate of the 
third term ($T_3$).
Therefore after including the proper prefactor in Eq. (\ref{akak}),
the total effect of non-condensate particle can be
obtained by adding them together to be
\be
\langle a^\dagger_\bfk a^{}_\bfk\rangle_{\bfk\neq 0} &=& (n_0+1)T_1(\bfk)+
n_0T_2(\bfk)
\nonumber\\
&&+\sqrt{n_0(n_0+1)}(T_3(\bfk)+T_4(\bfk))
\nonumber\\
&\equiv&\vec{b}_\bfk^\dagger\cdot {\cal T}(\bfk)\cdot\vec{b}_\bfk,
\ee
where $\vec{b}_\bfk\equiv 
[b^{}_{1,\bfk},b^{}_{2,\bfk},b^\dagger_{1,-\bfk},
b^\dagger_{2,-\bfk}]^T$ and ${\cal T}(\bfk)$ is the result of the final $4\times 4$
matrix ({\it not} the same as the $3\times 3$ transform matrix, ${\cal U}$,
defined in Eq. (\ref{cal_U_def})).

In the final step, we have to transform the operator of $b_{1/2,\pm\bfk}$ 
to $\beta_{s/m,\pm\bfk}$ (as shown in Eq. (\ref{b-beta})) to calculate
the expectation value. The obtain momentum distribution from 
the noncondensate particles can therefore be written to be
\be
\langle a^\dagger_\bfk a^{}_\bfk\rangle_{\bfk\neq 0}
&=&\vec{b}_\bfk^\dagger\cdot {\cal T}(\bfk)\cdot\vec{b}_\bfk
\nonumber\\
&=&\vec{\beta}^\dagger_\bfk\cdot {\cal M}(\bfk)^\dagger
\cdot {\cal T}(\bfk)\cdot {\cal M}(\bfk) \vec{\beta}_\bfk
\nonumber\\
&\equiv&\vec{\beta}^\dagger_\bfk \cdot {\cal S}(\bfk) \cdot\vec{\beta}_\bfk,
\label{akak_SF0}
\ee
where the new matrix ${\cal S}(\bfk)\equiv {\cal M}(\bfk)^\dagger
\cdot {\cal T}(\bfk)\cdot {\cal M}(\bfk)$. Using the fact that
$\langle\beta_{s/m,\bfk}^\dagger\beta_{s/m,\bfk}\rangle
=f_B(\epsilon_{s/m,\bfk})$ for the Bose-Einstein distribution of 
particle/hole excitation.
Therefore, the final result of the noncondensate particle 
momentum distribution can be expressed by the matrix elements
of matrix ${\cal S}(\bfk)$:
\be
\langle a^\dagger_\bfk a^{}_\bfk\rangle_{\bfk\neq 0}
&=&(S_{11}(\bfk)^2+S_{33}(\bfk)^2)f(\epsilon_{s,\bfk})
\nonumber\\
&&+(S_{22}(\bfk)^2+S_{44}(\bfk)^2)f(\epsilon_{m,\bfk})
\nonumber\\
&&+S_{33}(\bfk)^2+S_{44}(\bfk)^2,
\label{akak_SF}
\ee
where the last two terms are the contribution of quantum deplated
particles.

\section{Finite temperature momentum 
distribution in Mott insulator phase}
\label{appendix_MI}

\subsection{General expression in three-state model}
\label{appendix_n_k_MI_general}

In the Mott insulator phase, the meanfield ground state has
one particle persite. Therefore it is important to separate the
contribution of onsite density operator, $a^\dagger_\bfR a^{}_\bfR$,
from other non-local density operators. In order words, when calculating
the momentum distribution, we have
\wbe
\be
\langle a_\bfk^\dagger a_{\bfk}\rangle&=&\frac{1}{L^3}\sum_{\bfR_1,\bfR_2}
a^\dagger_{\bfR_1}a^{}_{\bfR_2}\,e^{i\bfk\cdot(\bfR_1-\bfR_2)}
=\frac{1}{L^3}\sum_{\bfR}
\langle a^\dagger_{\bfR}a^{}_{\bfR}\rangle
+\frac{1}{L^3}\sum_{\bfR_1\neq\bfR_2}
\langle a^\dagger_{\bfR_1}a^{}_{\bfR_2}\rangle
\,e^{i\bfk\cdot(\bfR_1-\bfR_2)}
\nonumber\\
&=&\frac{1}{L^3}\sum_{\bfR}
\langle\left[(n_0+1)t^\dagger_{1,\bfR}t^{}_{1,\bfR}
+n_0t^\dagger_{0,\bfR}t^{}_{0,\bfR}
+(n_0-1)t^\dagger_{-1,\bfR}t^{}_{-1,\bfR}\right]\rangle
\nonumber\\
&&-\frac{1}{L^3}\sum_{\bfR}
\langle\left[\sqrt{n_0+1}t^\dagger_{1,\bfR}t^{}_{0,\bfR}
+\sqrt{n_0}t^\dagger_{0,\bfR}t^{}_{-1,\bfR}\right]
\left[\sqrt{n_0+1}t^\dagger_{0,\bfR}t^{}_{1,\bfR}
+\sqrt{n_0}t^\dagger_{-1,\bfR}t^{}_{0,\bfR}\right]\rangle
\nonumber\\
&&+\frac{1}{L^3}\sum_{\bfR_1,\bfR_2}
\langle\left[\sqrt{n_0+1}t^\dagger_{1,\bfR_1}t^{}_{0,\bfR_1}
+\sqrt{n_0}t^\dagger_{0,\bfR_1}t^{}_{-1,\bfR_1}\right]
\left[\sqrt{n_0+1}t^\dagger_{0,\bfR_2}t^{}_{1,\bfR_2}
+\sqrt{n_0}t^\dagger_{-1,\bfR_2}t^{}_{0,\bfR_2}\right]
\rangle e^{i\bfk\cdot(\bfR_1-\bfR_2)},
\nonumber\\
\label{akak_MI}
\ee
\wee
where in the last line, for the convenience of later
Fourier tranform into momentum space, 
we have added the term of $\bfR_1=\bfR_2$ 
and substracted it in the second term.

Before transforming to the momentum space representation
for the calculation of quasi-particle distribution, we can simplify
above results further by using the number constrain,
$\sum_{\alpha=-1}^1 t^\dagger_{n_0+\alpha,\bfR}
t^{}_{n_0+\alpha,\bfR}=1$ to replace the field operator,
$t_{0,\bfR}$ and $t^\dagger_{0,\bfR}$ to the
quadratic order of small fluctuation,
After some straightforward calculation, 
it is easy to show that
\wbe
\be
\langle a_\bfk^\dagger a_{\bfk}\rangle 
&=&n_0-\frac{1}{L^3}\sum_{\bfk}
\left[(n_0+1) \langle t^\dagger_{1,\bfk}t^{}_{1,\bfk}\rangle
+n_0\langle t^{}_{-1,\bfk}t^{\dagger}_{-1,\bfk}\rangle+
\sqrt{n_0(n_0+1)}\left(\langle
t^{}_{-1,-\bfk}t^{}_{1,\bfk}\rangle
+\langle t^\dagger_{1,\bfk}t^{\dagger}_{-1,-\bfk}\rangle\right)\right]
\nonumber\\
&&+(n_0+1)\langle t^\dagger_{1,\bfk}t^{}_{1,\bfk}\rangle
+n_0\langle t^{}_{-1,\bfk}t^{\dagger}_{-1,\bfk}\rangle
+\sqrt{n_0(n_0+1)}
\left[\langle t^{}_{-1,-\bfk}
t^{}_{1,\bfk}\rangle
+\langle t^{\dagger}_{1,\bfk}t^{\dagger}_{-1,-\bfk}\rangle\right],
\ee
\wee
where in the last line we apply Fourier transofrm,
$t_{\alpha,\bfk}=\frac{1}{L^{3/2}}\sum_\bfR 
t_{\alpha,\bfR}e^{-i\bfk\cdot\bfR}$, to the momentum
space representation. Note that the first line of the last equation
is independent of momentum $\bfk$, and therefore can be
understood as a constant shift of the momentum distribution.

We now calculate the expectation value for each term 
by transforming to the quasi-particle, $\beta_\bfk$, field via Eq. 
(\ref{t-beta}) and neglecting those
particle non-conserving terms (for example, 
$\langle\beta_{p,\bfk}\beta_{p,\bfk}\rangle=\langle 
\beta^\dagger_{h,\bfk}\beta^\dagger_{h,\bfk}\rangle
=\langle\beta^\dagger_{p,\bfk}\beta^{}_{h,\bfk}\rangle=0$). 
We have
\wbe
\be
\langle t^\dagger_{1,\bfk}t^{}_{1,\bfk}\rangle 
&=&A(\bfk)^2\langle\beta^\dagger_{p,-\bfk}
\beta^{}_{p,-\bfk}\rangle
+B(\bfk)^2\langle\beta^{}_{h,\bfk}\beta^{\dagger}_{h,\bfk}
\rangle=A(\bfk)^2f_B(\epsilon_{p,\bfk})+B(\bfk)^2
(1+f_B(\epsilon_{h,\bfk})),
\\
\langle t^{}_{-1,\bfk}t^{\dagger}_{-1,\bfk}\rangle 
&=&B(\bfk)^2f_B(\epsilon_{p,\bfk})+A(\bfk)^2
(1+f_B(\epsilon_{h,\bfk})),
\\
\langle t^{\dagger}_{1,\bfk}t^{\dagger}_{-1,-\bfk}\rangle 
&=&\langle t^{}_{-1,-\bfk}t^{\dagger}_{1,\bfk}
\rangle^\ast 
=-A(\bfk)B(\bfk)f_B(\epsilon_{p,\bfk})-A(\bfk)B(\bfk)
(1+f_B(\epsilon_{h,\bfk})),
\ee
\wee
where $f_B(\epsilon_{p/h,\bfk})=\langle\beta^\dagger_{p/h,-\bfk}
\beta^{}_{p/h,-\bfk}\rangle$ is the Bose-Einstein distribution of
particle/hole excitations. 

Therefore we can obtain the final expression of the 
momentum distribution, $n(\bfk)\equiv\langle a^\dagger_\bfk 
a^{}_\bfk\rangle$, to be
\be
n(\bfk)&=&-\delta n
+f_B(\epsilon_{p,\bfk})\left(\sqrt{n_0+1}A(\bfk)-\sqrt{n_0}B(\bfk)\right)^2
\nonumber\\
&&+(1+f_B(\epsilon_{h,\bfk}))
\left(\sqrt{n_0+1}B(\bfk)-\sqrt{n_0}A(\bfk)\right)^2,
\nonumber\\
\label{n_k_MI}
\ee
where the density shift, $\delta n$ is defined by
\be
\delta n &\equiv &\frac{1}{L^3}\sum_{\bfk}
\left[\left(\sqrt{n_0+1} A(\bfk)-\sqrt{n_0}B(\bfk)\right)^2
f_B(\epsilon_{p,\bfk})\right.
\nonumber\\
&&\left.+\left(\sqrt{n_0+1} B(\bfk)
-\sqrt{n_0}A(\bfk)\right)(1+f_B(\epsilon_{h,\bfk}))
\right]-n_0
\nonumber\\
\label{delta_n}
\ee
to ensure the conservation of total number of particles.

\subsection{Analytical calculation of chemical potential}

In order to evalulate the momentum distribution in the normal
state regime, the first thing
is to determine the chemical potential 
by fixing the total number of particles. We can use the general 
expression of momentum distribution, i.e. 
$\langle a^\dagger_\bfR a^{}_\bfR\rangle$ in Eq. (\ref{akak_MI}),
and obtain the total number of particles to be
\be
N&=&\sum_\bfk{}'\langle a^\dagger_\bfk a^{}_\bfk\rangle
=\sum_\bfR\langle a^\dagger_\bfR a^{}_\bfR\rangle
\nonumber\\
&=&L^3n_0+\sum_\bfR\langle t^\dagger_{1,\bfR}t^{}_{1,\bfR}
-t^\dagger_{-1,\bfR}t^{}_{-1,\bfR}\rangle
\nonumber\\
&=&N+\sum_\bfk{}'\langle t^\dagger_{1,\bfk}t^{}_{1,\bfk}
-t^\dagger_{-1,\bfk}t^{}_{-1,\bfk}\rangle.
\ee
Now if we tranfer the $t-$operator to the quasi-particle
operators, $\beta_{p/h,\bfk}$, in the MI state via Eq. (\ref{t-beta}),
the conservation of total number of particle can be expressed to be
\be
0&=&\sum_\bfk \left(A(\bfk)^2-B(\bfk)^2\right)
\left[f_B(\epsilon_{p,\bfk})-f_B(\epsilon_{h,\bfk})\right]
\nonumber\\
&=&\sum_\bfk 
\left[f_B(\epsilon_{p,\bfk})-f_B(\epsilon_{h,\bfk})\right],
\label{mu_eq}
\ee
which is nothing but the difference of particle excitations and hole
excitations. Below we will investigate
analytically the chemical potential in different parameter regimes of
interest.

\subsubsection{In the low temperature limit, $U\gg T,J$}

We first consider the situation when $U\gg T,J$, so that
we can use large $U$ expansion
for the particle and hole excitation spectrum and obtain 
\be
\epsilon_{p,\bfk}&=&\frac{U}{2}-(\delta\mu+(n_0+1)\epsilon_0(\bfk))
\nonumber\\
&&-\frac{n_0(n_0+1)\epsilon_0(\bfk)^2}{U}+{\cal O}((J/U)^{2})
\\
\epsilon_{h,\bfk}&=&\frac{U}{2}-(-\delta\mu+n_0\epsilon_0(\bfk))
\nonumber\\
&&-\frac{n_0(n_0+1)\epsilon_0(\bfk)^2}{U}+{\cal O}((J/U)^{2}).
\ee
Since $U\gg T$, we can expand the Bose-Eistein distribution 
fuction as 
\be
f_B(\epsilon(\bfk))&=&\frac{1}{e^{\epsilon(\bfk)/T}-1}
\nonumber\\
&=&\frac{1}{e^{(U/2-\delta\epsilon(\bfk)\mp\delta\mu)/T}-1}
\nonumber\\
&=&e^{-(U/2-\delta\epsilon(\bfk)\mp\delta\mu)/T}
+{\cal O}(e^{-U/T}),
\label{f_ph_large_U}
\ee
where $\delta\epsilon(\bfk)\equiv(n_0+1)\epsilon_0(\bfk)$ for particle 
excitation and
$\delta\epsilon(\bfk)\equiv n_0\epsilon_0(\bfk)$ for hole excitation.
Now using the fact that $\int_{-\pi}^\pi\frac{dk}{2\pi}
e^{x\cos(k)}=I_0(x)$, where $I_0(x)$ is the modified Bessel
function of the first kind, we can integrate out the momentum
integral of Eq. (\ref{mu_eq}) and obtain
\be
\frac{1}{L^3}\sum_\bfk{}'f_B(\epsilon_p(\bfk))
=e^{-U/2T}e^{\delta\mu/T}
\left[I_0\left(\frac{2(n_0+1)J}{T}\right)\right]^3
\nonumber\\
\ee
and
\be
\frac{1}{L^3}\sum_\bfk{}'f_B(\epsilon_h(\bfk))
=e^{-U/2T}e^{-\delta\mu/T}
\left[I_0\left(\frac{2n_0J}{T}\right)\right]^3.
\ee
Here we have neglected the higher order terms, which is proportional
to $e^{-U/T}\ll 1$.
As a result, the chemical potential can be calculated to be
(from Eq. (\ref{mu_eq}))
\be
e^{\delta\mu/T}&=&\left(\frac{I_0\left(\frac{2n_0J}{T}\right)}
{I_0\left(\frac{2(n_0+1)J}{T}\right)}\right)^{3/2},
\label{mu_MI_lowT}
\ee
which is good for all the temperature range if only $U\gg J,T$.
If consider intermediate temperature regime, $U\gg T\gg J$, 
which is the most relevant regime in the present experiment, 
the leading order result gives
\be
\frac{\delta\mu}{J}=-\frac{z}{4}(1+2n_0)\frac{J}{T}
\ee
where $z=6$ in 3D and $z=4$ for 2D systems.

On the other hand, in the limit of small $T$ (i.e. $T\ll J$), 
we use $I_0(x)=\sqrt{\frac{\pi}{2 x}}e^x$ and obtain
\be
e^{\delta\mu/T}&=&\left(\frac{n_0+1}{n_0}\right)^{3/4}e^{-3J/T}
\ee
or 
\be
\delta\mu&=&-\frac{zJ}{2}+\frac{8T}{z}\log\left(\frac{n_0+1}{n_0}\right)
\ee
This is consistent with the results at $T=0$.

\subsubsection{In high temperature interaction limit, $T\sim U\gg J$}

In the limit of high temperature, we can expand the excitation
modes and Bose-einstein distribution to the leading order terms
of small $J/T$ and $J/U$. we have 
\wbe
\be
\frac{1}{e^{\epsilon(\bfk)/T}-1}
&=&\frac{1}{e^{U/2T}e^{-(\delta\epsilon(\bfk)\pm\delta\mu)/T}-1}
=\frac{1}{e^{U/2T}-1}
+\frac{e^{U/2T}}{(e^{U/2T}-1)^2}\frac{\delta\epsilon(\bfk)\pm\delta\mu}{T}
\nonumber\\
&&+\frac{1}{(e^{U/2T}-1)^2}\left(\frac{e^{U/T}}{e^{U/2T}-1}-
\frac{1}{2}e^{U/2T}\right)
\left(\frac{\delta\epsilon(\bfk)\pm\delta\mu}{T}\right)^2+\cdots
\ee
Therefore within the same order, the total number of particle and hole 
excitations shown in Eq. (\ref{mu_eq}) can be calculated to be
(note that $\epsilon_0(\bfk)=2J\sum_{\alpha}\cos(k_\alpha a)$ for 3D optical 
lattice.)
\be
\frac{1}{\Omega}\sum_\bfk{}'f_B(\epsilon_p(\bfk))
&=&\frac{1}{e^{U/2T}-1}
+\frac{e^{U/2T}}{(e^{U/2T}-1)^2}\frac{\delta\mu}{T}
+\frac{1}{(e^{U/2T}-1)^2}\left(\frac{e^{U/T}}{e^{U/2T}-1}-
\frac{1}{2}e^{U/2T}\right)
\left(\frac{\delta\mu^2}{T^2}+z(n_0+1)^2\frac{J^2}{T^2}\right)
\nonumber\\
\\
\frac{1}{\Omega}\sum_\bfk{}'f_B(\epsilon_h(\bfk))
&=&\frac{1}{e^{U/2T}-1}
+\frac{e^{U/2T}}{(e^{U/2T}-1)^2}\frac{-\delta\mu}{T}
+\frac{1}{(e^{U/2T}-1)^2}\left(\frac{e^{U/T}}{e^{U/2T}-1}-
\frac{1}{2}e^{U/2T}\right)
\left(\frac{\delta\mu^2}{T^2}+zn_0^2\frac{J^2}{T^2}\right).
\ee
\wee
The chemical potential then can be determined by the following 
colse form to the second order of $J/T$ 
\be
\frac{\delta\mu}{T}&=&\frac{-z(2n_0+1)}{4}\coth\left(\frac{U}{4T}\right)
\left(\frac{J}{T}\right)^2.
\label{mu_high_T}
\ee
in the high temperature and strong interaction limit, $U\sim T\gg J$.

\subsection{Analytical calculation of momentum distribution}

\subsubsection{In the low temperature limit, $U\gg T,J$}

In order to study the momentum distribution shown in 
Eq. (\ref{n_k_MI}), we can start from the following  
leading order expansion in the strong interaction limit 
($U\gg T,J$) by using Eq. (\ref{D}): 
$(\sqrt{n_0+1}B(\bfk)-\sqrt{n_0}A(\bfk))^2
\sim n_0$, and
$(\sqrt{n_0+1}A(\bfk)-\sqrt{n_0}B(\bfk))^2\sim 1+n_0$.
Now, after including the large $U$ expansion of particle-hole excitation
shown in Eq. (\ref{f_ph_large_U}), the density shift can be 
expressed to the leading order to be
\be
\delta n&\equiv &\frac{1}{L^3}\sum_\bfk{}'\left[
(n_0+1) e^{-U/2T}
e^{(n_0+1)\epsilon_0(\bfk)/T}e^{\delta\mu/T}\right.
\nonumber\\
&&\left.+n_0\left(1+e^{-U/2T}
e^{n_0\epsilon_0(\bfk)/T}e^{-\delta\mu/T}\right)\right]-n_0
\nonumber\\
&=&(2n_0+1)e^{-U/2T}I_0\left(\frac{2n_0J}{T}\right)^{3/2}
I_0\left(\frac{2(n_0+1)J}{T}\right)^{3/2}.
\nonumber\\
\label{delta_n_final}
\ee
The momentum distribution then becomes
\wbe
\be
n(\bfk)&=&
(\sqrt{n_0+1}A(\bfk)-\sqrt{n_0}B(\bfk))^2f_B(\epsilon_{p,\bfk})
+(\sqrt{n_0+1}B(\bfk)-\sqrt{n_0}A(\bfk))^2(1+f_B(\epsilon_{h,\bfk}))
-\delta n
\nonumber\\
&=&n_0+2n_0(1+n_0)\frac{\epsilon_0(\bfk)}{U}
+(1+n_0)\left(\frac{I_0(\frac{2n_0J}{T})}
{I_0(\frac{2(n_0+1)J}{T})}\right)^{3/2}e^{-U/2T}
e^{(n_0+1)\epsilon_0(\bfk)/T}
\nonumber\\
&&+n_0\left(\frac{I_0(\frac{2(n_0+1)J}{T})}
{I_0(\frac{2n_0J}{T})}\right)^{3/2}
e^{-U/2T}e^{n_0\epsilon_0(\bfk)/T}
-(2n_0+1)e^{-U/2T}I_0\left(\frac{2n_0J}{T}\right)^{3/2}
I_0\left(\frac{2(n_0+1)J}{T}\right)^{3/2},
\label{nk_MI_lowT}
\ee
\wee
to the leading order of large $U$, i.e. the next higher order
is proportional to $(J/U)\,e^{-U/2T}$. 
It is easy to show that the 
total number of particles is conserved, i.e. $\frac{1}{L^3}
\sum_\bfk{}' n(\bfk)=n_0$, and therefore above expression is
a {\it close} form for the momentum distribution in large $U$ limit.
It is also straightforward to calculate the
higher order terms but we will not show them here.
From the momentum distribution in the optical lattice, we can also
integrate out the $z$-component of the lattice momentum and obtain
\be
n_\perp(\bfk_\perp)&=&\frac{1}{L}\sum_{k_z}{}'n(\bfk)
\nonumber\\
&=&n_0-\delta n+2n_0(1+n_0)\frac{\epsilon_0(\bfk_\perp)}{U}
\nonumber\\
&&+G_1e^{(n_0+1)\epsilon_0(\bfk_\perp)/T}
+G_2 e^{n_0\epsilon_0(\bfk_\perp)/T},
\nonumber\\
\label{N_perp_MI_lowT}
\ee
where $\epsilon_0(\bfk_\perp)=2J(\cos(k_xa)+\cos(k_ya))$ and
\be
G_1&=&(1+n_0)e^{-U/2T}\frac{I_0(\frac{2n_0J}{T})^{3/2}}
{I_0(\frac{2(n_0+1)J}{T})^{1/2}}
\nonumber\\
G_2&=&n_0e^{-U/2T}\frac{I_0(\frac{2(n_0+1)J}{T})^{3/2}}
{I_0(\frac{2n_0J}{T})^{1/2}}
\label{NA_perp_MI_lowT}
\ee
%

\subsubsection{High temperature limit, $T\sim U\gg J$}

For the high temperature limit, we need
to expand the whole formular in terms of $J/T$ and $J/U$ at the same
time. To the leading order, we thus have
\wbe
\be
n(\bfk)&=&(\sqrt{n_0+1}A(\bfk)-\sqrt{n_0}B(\bfk))^2
f_B(\epsilon_{p,\bfk})
+(\sqrt{n_0+1}B(\bfk)-\sqrt{n_0}A(\bfk))^2(1+f_B(\epsilon_{h,\bfk}))
-\delta n
\nonumber\\
&\sim&\left(1+n_0
+2n_0(1+n_0)\frac{\epsilon_0(\bfk)}{U}\right)
\frac{1}{e^{U/2T}-1}\left(1
+\frac{e^{U/2T}}{e^{U/2T}-1}\frac{(n_0+1)\epsilon_0(\bfk)+\delta\mu}{T}\right)
\nonumber\\
&&+\left(n_0+2n_0(1+n_0)\frac{\epsilon_0(\bfk)}{U}\right)
\left[1+\frac{1}{e^{U/2T}-1}\left(1
+\frac{e^{U/2T}}{e^{U/2T}-1}\frac{n_0\epsilon_0(\bfk)
-\delta\mu}{T}\right)\right]-\delta n
\nonumber\\
&\sim & n_0+2n_0(1+n_0)\frac{\epsilon_0(\bfk)}{U}
+\frac{4n_0(1+n_0)}{e^{U/2T}-1}
\frac{\epsilon_0(\bfk)}{U}
+\frac{(1+2n_0+2n_0^2)e^{U/2T}}{(e^{U/2T}-1)^2}
\frac{\epsilon_0(\bfk)}{T},
\label{nk_MI_highT}
\ee
\wee
where the correction of chemical potential is the second order
effect (Eq. (\ref{mu_high_T})) and has been neglected.
In the last line above, we also used the fact that $\delta n$ is determined
by requiring $\frac{1}{L^3}\sum_\bfk{}'n(\bfk)=n_0$ to the leading
order, and therefore obtained $\delta n=\frac{1+2n_0}{e^{U/2T}-1}$.
Therefore, after integration over the $k_z$ component, we have
\be
n_\perp({\bfk_\perp})&=&n_0+H_1 \epsilon_0(\bfk_\perp)/J,
\label{nk_perp_MI_highT}
\ee
where
\be
H_1&=&\frac{2n_0(1+n_0)J}{U}\left[1
+\frac{2}{e^{U/2T}-1}\right]
\nonumber\\
&&+\frac{(1+2n_0+2n_0^2)e^{U/2T}}{(e^{U/2T}-1)^2}
\frac{J}{T}.
\label{H_1}
\ee
%



\begin{thebibliography}{99}
\bibitem{interaction_TWA}
The interaction effects during the TOF expansion 
was discussed shortly in Refs. [\oncite{duan,finite_time}] below. 
But a quantitative study has not been systematically investigated
to the best of our knowledge.
\bibitem{Stanford}
Fitting with three Gaussian-type wavefunctions are also used
in recent experiments of Stanford's group. Private communication.
\end{thebibliography}
\end{document}